\documentclass[preprint,12pt,square,comma,number]{elsarticle}
\usepackage{numcompress}
\usepackage{amsmath}
\usepackage{color,txfonts}
\usepackage{amssymb}
\usepackage{graphicx}
\usepackage{amssymb}
\usepackage{verbatim}
\usepackage{booktabs}
\usepackage{subfig}
\usepackage[colorlinks,
            linkcolor=blue,
            anchorcolor=blue,
            citecolor=blue]{hyperref}
\begin{document}

\begin{frontmatter}

% \title{The Study of STK Alignment and Global Alignment of DAMPE Detector(v2)}
%\title{STK Alignment and Global Alignment of the DAMPE Detector}
\title{Study of the Global Alignment for the DAMPE Detector}


\author[1,2]{ Yu-Xin Cui}

\author[1]{ Peng-Xiong Ma}

\author[1,2]{ Guan-Wen Yuan}

\author[1]{ Chuan Yue \corref{cor1}}
\ead{yuechuan@pmo.ac.cn}

\author[1]{ Xiang Li \corref{cor1}}
\ead{xiangli@pmo.ac.cn}

\author[1]{ Shi-Jun Lei}

\author[1,2]{ Jian Wu}

\cortext[cor1]{Corresponding author}

\address[1]{Key Laboratory of Dark Matter and Space Astronomy, Purple Mountain Observatory, Chinese Academy of Sciences, Nanjing 210008, China}
\address[2]{School of Astronomy and Space Science, University of Science and Technology of China, Hefei 230026, China}

\begin{abstract}
The Dark Matter Particle Explorer (DAMPE) is designed as a high energy particle detector for probing cosmic-rays and $\gamma-$rays in a wide energy range. 
The trajectory of the incident particle is mainly measured by the Silicon-Tungsten tracKer-converter (STK) sub-detector, which heavily depends on the precise internal alignment correction as well as the accuracy of the global coordinate system.  
%The Silicon-Tungsten tracKer-converter (STK), one of the key sub-detector of the DAMPE instrument, is designed to measure the trajectory and charge of incident cosmic ray. 
% This work validates the potential issues of self-alignment through simulations, and the STK alignment result from the flight data is then used to align the BGO and complete the overall alignment of the detector, enabling the detector to satisfy the required objectives well.
In this work, we carried out a global alignment method to validate the potential displacement of these sub-detectors, and particularly demonstrated that the track reconstruction of STK can well satisfy the required objectives by means of comparing flight data and simulations.
\end{abstract}


\end{frontmatter}


\section{Introduction}\label{}
\begin{figure}[htb]\centering
\includegraphics[scale=0.57]{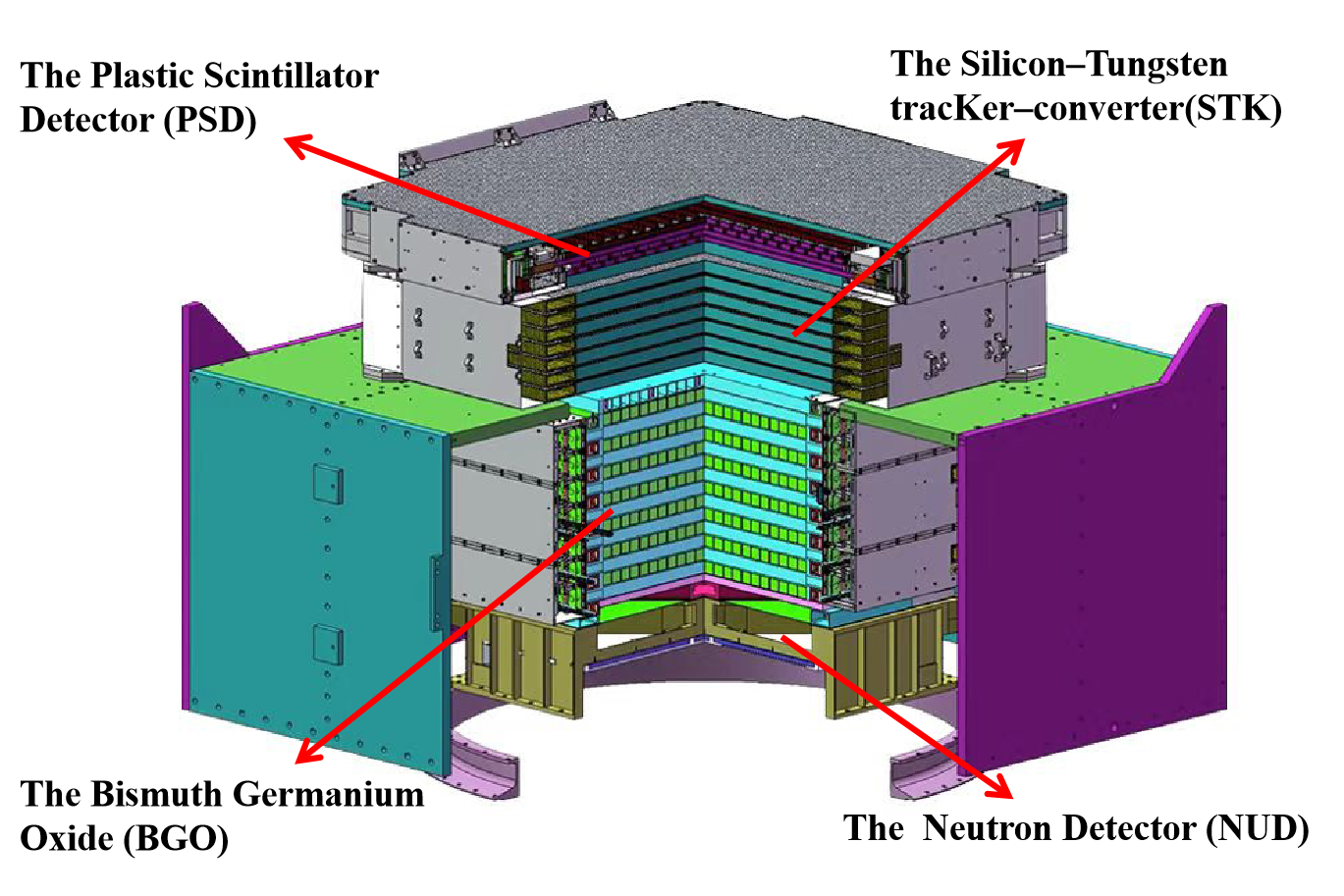}
\caption{A schematic of the DAMPE detector showing the four sub-detectors from up to down: the PSD, the STK, the BGO, and the NUD.}
\label{fig:dampe}
\end{figure}

The Dark Matter Particle Explorer (DAMPE) satellite, a space-borne experiment for high energy cosmic-ray and $\gamma$-ray observations~\cite{DAMPE,DAMPE:2017fbg}, was launched on December 17, 2015 and has been operating smoothly ever since~\cite{DAMPE:2019lxv}. The DAMPE payload is composed of four sub-detectors~\cite{DAMPE}, a Plastic Scintillator Detector (PSD), a Silicon–Tungsten tracKer–converter(STK), a Bismuth Germanium Oxide (BGO) imaging calorimeter, and a Neutron Detector (NUD), as shown schematically in Figure \ref{fig:dampe}.
The PSD measures the particle charge and serves as an anti-coincidence detector for $\gamma-$ray~\cite{Yu:2017dpa, Dong:2018qof}. The STK measures the trajectory of the incident charged particle and converts $\gamma-$ray into electron and positron pairs~\cite{dampe:stk, stk_temp}. The BGO imaging calorimeter~\cite{ZhangZY:2016}, which is of about 31 radiation lengths, measures the particle energy with a high resolution~\cite{YueC:2017}. The NUD enhances the capability to discriminate electron/positrons from protons in high energy range~\cite{Huang:2020skz}.

DAMPE is benefited from a very high energy resolution and an excellent trajectory accuracy, which, however, might be affected by the potential displacement of the sensitive units in the detector. Thus, the alignment of each sub-detector and their global performance are essential for the measurement. For DAMPE, although alignments of the PSD and the STK have been considered separately in Ref.~\cite{psd_ali,stk_ali}, a global alignment still needs to be further studied. Especially, previous works ignore the displacement, rotation, and scaling of the STK as a whole in the detector. An overall cross check associating all the sub-detectors is thus essentially required for evaluating these effects.

%Moreover, as the position resolution of STK is much better than the one of BGO, previous works ignores the effects of displacement, rotation and scaling of sub-detectors. \mkred{This point to the displacement between BGO and STK?  done} Thus, an overall cross check by associating all sub-detectors is urgently required for evaluating these effects.

% Nevertheless, position resolution and the size of the units of the other sub-detectors are much weaker than that of STK thanks to its tiny and sensitive unit. The PSD and BGO cannot be substituted directly into the global alignment procedure. 
% As a result, the global alignment is accomplished in two steps: first, the STK's self-alignment; and secondly, using the STK's alignment tracks to align the PSD and BGO. 
% On the one hand, global alignment can maintain each sub-detector's relative position and improve measurement accuracy. On the other hand, the previously indicated STK self-alignment could result in the effects of displacement, rotation, and scaling~\cite{rotation}. Global alignment could evaluate whether the STK alignment parameters are within a reasonable range.

\begin{figure}[htbp]
\centering
\includegraphics[scale=0.34]{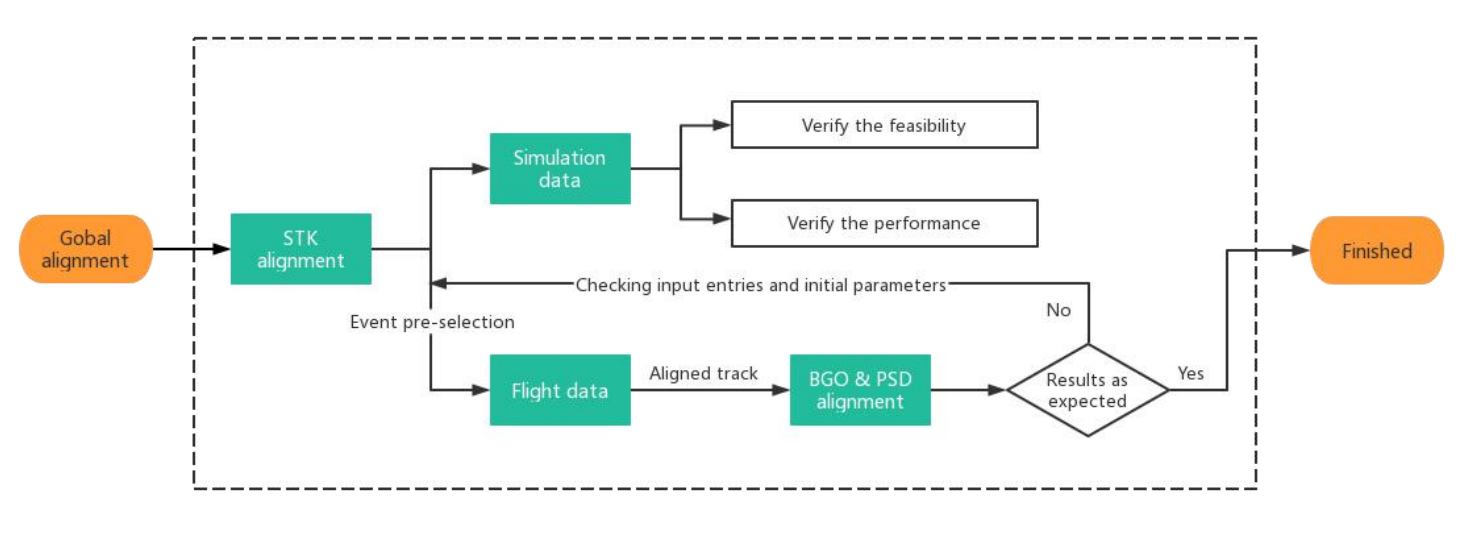}
\caption{General scheme of global alignment.}
\label{Flowchart}
\end{figure}

In this work, the self-alignment of the STK is carried out firstly, as the size of its sensitive units is much smaller than that of the PSD and the BGO. Then, for the global alignment in a second step, we use the tracks already aligned for the STK to calibrate the overall displacement of the sTK relative to the other sub-detectors.
This method not only improves the measurement accuracy, but also maintains the correct relative position of different sub-detectors. As the self-alignment of STK could induce the effects of displacement, rotation and scaling~\cite{rotation}, the global alignment helps us evaluate whether the alignment parameters are within reasonable range. A schematic view of the entire alignment procedure is shown in Figure~\ref{Flowchart}.

The rest of this paper is organized as the following. In Sec.(\ref{sec.stk_ali}), we provide a quick overview of the STK, as well as the principle and method of the STK self-alignment, including the use of simulation data to identify potential issues. Then, in Sec.(\ref{gloabl_alignment}), we introduce the global alignment procedure. Finally, We conclude this work in Sec.(\ref{sec:conclusion}). 

\section{STK alignment}\label{sec.stk_ali}
\subsection{STK Detector}
The STK is used to measure the trajectory of incident primary particle, which is essential for the particle charge identification and the acceptance estimation. It is made of 6 tracking double layers, and each consists of two layers of single-sided silicon strip detectors, which are orthogonal with each other and views perpendicularly to the pointing direction of the apparatus. The STK has 768 silicon micro-strip detectors (sensors) that are single-sided AC-coupled and have the same geometry as the AGILE.~\cite{agile}. Each sensor in STK is 9.5 cm $\times $ 9.5 cm in size, 320 $\rm \mu m$ thick, and segmented into 768 strips with a 121 $\rm \mu m$ pitch \cite{dampe:stk}. Only every other strip is read out, but since analogue readout is used, the position resolution is better than 80 $\rm \mu m$ for most incident angles, thanks to the charge division of floating strips.

\begin{comment}
\begin{figure}[htb]\centering
\includegraphics[scale=0.6]{fig/stk2.png}
\caption{Schematic of the STK, it is made of 6 tracking double layers, and each consists of two layers of single-sided silicon strip detectors, which are orthogonal with each other and views perpendicularly to the pointing direction of the apparatus.}
\label{fig:stk}
\end{figure}
\end{comment}

\subsection{Methodology of STK alignment}
Alignment analysis with tracks uses the fact that the hit positions and the measured trajectory impact points of a track are systematically displaced if the module position is not known correctly~\cite{2009cms}.

The method used to determine the alignment parameters is based on the minimization of the total $\chi^{2}$ of tracks in the alignment data samples, and the $\chi^{2}$ can be written as:

\begin{equation}\label{chi2}
\chi ^{2} = \sum_{i}^{\rm tracks} \sum_{j}^{\rm points} \left [\frac{\left (  x_{ij}^{\rm fit}-x_{ij}^{\rm hit} \right ) ^{2} }{\sigma_{x}^{2}}+  \frac{\left ( y_{ij}^{\rm fit}-y_{ij}^{\rm hit} \right ) ^{2}}{\sigma_{y}^{2}} \right ], 
\end{equation}
where $x_{ij}^{\rm fit}$,$y_{ij}^{\rm fit}$ and $x_{ij}^{\rm hit}$,$y_{ij}^{\rm hit}$ are the positions with and without aligned, subscript $i,j$ are tracks and points respectively, $\sigma_{x}$ and $\sigma_{y}$ are their uncertainties.

%To ensure the reliability of the tracks, only the tracks with 12 hit points and the residuals of less than 0.4 mm are retained. The residual is defined as the difference between the measured coordinate and the projection of the track in the $i$-th plane where the track is reconstructed without considering the $i$-th plane. In order to prevent the influence of large Coulomb scattering and measurement errors on the overall results, it is necessary to first eliminate the points with large deviations and low energy, which is avoided by means of the cut that BGO energy should be higher than 15 GeV. 

Only the tracks with 12 hit points and residuals of less than 0.4 mm are kept in order to guarantee the dependability of tracks. The residual is defined as the difference between the measured coordinate and the projection of the track in the $i$-th plane where the track is reconstructed without considering the $i$-th plane. In order to prevent the influence of large Coulomb scattering and measurement errors on the overall results, it is necessary to first eliminate the points with large deviations and low energy, applied by means of event cut that the BGO energy should be higher than 15 GeV. 

For track reconstruction, the local (sensor) coordinate system and the global coordinate system are used. The global system is centered at the midpoint of the first layer of the BGO, whereas the local system is centered at the midpoint of the corresponding sensor. The hit coordinates obtained by the detector are transformed from the local system as follows:

% \begin{equation}\label{coordinate_matrix_simple}
% \begin{bmatrix}
%  x\\  y\\ z
% \end{bmatrix}^{\rm global}  =\begin{bmatrix}
%  x\\ y\\ z
% \end{bmatrix}^{\rm local}    +\begin{bmatrix}
%  x_{c}\\ y_{c}\\ z_{c}
% \end{bmatrix},
% \end{equation}

%This is, nevertheless, a perfect condition, the actual hit coordinates should be written as:
% But Eq.~\eqref{coordinate_matrix_simple} is an ideal situation, the precise hit coordinates should be expressed as:

\begin{equation}\label{coordinate_matrix}
\begin{bmatrix}
 x\\  y\\ z
\end{bmatrix}^{\rm global}  =\mathbf{R}\cdot \begin{bmatrix}
 x\\ y\\ z
\end{bmatrix}^{\rm local}  +
\begin{bmatrix}
\Delta x\\ \Delta y\\ \Delta z
\end{bmatrix} + \quad \begin{bmatrix}
 x_{c}\\ y_{c}\\ z_{c}
\end{bmatrix},
\end{equation}
where $(x,y,z)^{\rm global}$ and $(x,y,z)^{\rm local}$ represent the coordinates of hit in the global and local systems, respectively. $(\Delta x, \Delta y, \Delta z)$ and $(x_{c},y_{c},z_{c})$ are offsets alignment parameters and position of the sensor center in global coordinates.
And the rotation matrix $\mathbf{R}$ includes three components $R_{\alpha}, R_{\beta}, R_{\gamma}$, which can be expressed in matrix form:
\begin{equation}\label{matrix}
\mathbf{R}=\left(\begin{array}{ccc}
\cos \alpha & \sin \alpha & 0 \\
-\sin \alpha & \cos \alpha & 0 \\
0 & 0 & 1\end{array}\right) \\
\left(\begin{array}{ccc}
\cos \beta & 0 & -\sin \beta \\
0 & 1 & 0 \\
\sin \beta & 0 & \cos \beta
\end{array}\right) \\
\left(\begin{array}{ccc}
1 & 0 & 0 \\
0 & \cos \gamma & \sin \gamma \\
0 & -\sin \gamma & \cos \gamma
\end{array}\right).
\end{equation}

Because the STK's hardware architecture prevents it from deflecting at a considerable angle. Therefore, Eq.~\eqref{coordinate_matrix} could be approximately transformed into the following form~\cite{Karimaki:2003bd}:

\begin{equation}
\begin{cases}
x^{\rm global}=x_{c}+x^{\rm local} + \Delta x - y^{\rm local}\cdot\Delta\gamma+ (\Delta z + x^{\rm local}\cdot\Delta\beta  + y^{\rm local}\cdot\Delta\alpha)\cdot tan\psi , \\
y^{\rm global}=y_{c}+y^{\rm local} + \Delta y + x^{\rm local}\cdot\Delta\gamma+ (\Delta z + x^{\rm local}\cdot\Delta\beta  + y^{\rm local}\cdot\Delta\alpha)\cdot tan\vartheta , \\
z^{\rm global}=z_{c}+z^{\rm local} + \Delta z + x^{\rm local}\cdot\Delta\beta  + y^{\rm local}\cdot\Delta\alpha ,
\end{cases}
\end{equation}
where $\Delta\alpha$, $\Delta\beta$, $\Delta\gamma $ correspond to the rotation parameters along X,Y and Z plane, respectively. The quantity $ \psi $  is the angle between the track and the YZ$-$plane and $ \vartheta $ is the angle between the track and the XZ$-$plane,  $ \delta\cdot tan\psi $ and  $ \delta\cdot tan\vartheta $ are higher-order terms that can be neglected.

The alignment correction coefficients include three position parameters ($\Delta x$, $\Delta y$, $\Delta z$) and orientation parameters ($\Delta \alpha$, $\Delta \beta$, $\Delta\gamma$). Due to the use of a single-sided silicon micro-strip, a track to obtain a total of 12 points in the x and y planes of 6 points each, first by these points to fit a straight line. The fitted linear parameters, sensor location, and measurement points constitute a matrix $\mathbf{A}\mathbf{x} = \mathbf{b}$~\cite{axb}, with the total number of entries represented by N, then  x and b are one-dimensional vectors of size 4608 and  $ N\times 12 $, respectively. $ \mathbf{A} $ is a $(N\times 12) \times 4608$ matrix.

A good balance between the number of entries and the time period must be set, as too few instances will lead to overfitting and too many instances will need to be picked for a long period of time, during which the STK may have new displacements resulting in mistakes. Because such a matrix is sparse, with the majority of its components being zero, the sparse matrix solver BiCGSTAB (Bi-conjugate gradient stabilized technique), which comes with Eigen3~\cite{eigen}, can solve it in roughly 3 seconds for 100,000 entries. Following that, the solution's alignment parameters are corrected for the hit points, the track is re-fitted with the corrected points, and the procedure is continued until the Eq.~\eqref{chi2} converges to a specific level of precision.

% \section{STK self-alignment simulation}
\subsection{Pseudo-simulation}

%The precise position of each sensor cannot be determined from flight data, instead, the alignment algorithm must be confirmed by simulation. Because it is difficult to precisely simulate the performance of the STK interval readout using GEANT4\cite{GEANT4:2002zbu}, a numerical simulation approach is used to acquire the simulated coordinates more quickly and simply. Before numerical simulation, the way to characterize the rotation of a rigid body, such as Euler angle, Quaternions, Rotation matrix, and so on, must be determined. In this case, DCM(Direction Cosine Matrix)\cite{dcm}, a type of Rotation matrix, is used to simulate, according to sec.2, the simulation method is separated into three parts.

Flight data cannot determine the precise position of each sensor; instead, the alignment algorithm must be validated through simulation. Because precisely simulating the performance of the STK interval readout with GEANT4~\cite{GEANT4:2002zbu} is difficult, a numerical simulation approach is used to obtain the simulated coordinates more quickly and simply. Before numerical simulation, the way to characterize the rotation of a rigid body must be determined, such as the Euler angles, the quaternions, or the rotation matrix. The Direction Cosine Matrix (DCM)~\cite{dcm} is used in our simulation, which is a powerful tool in dealing rotation matrix. %the simulation method is separated into three parts.
There are three steps in our simulation:

\begin{itemize}
\item[a.] Give a track parameters as well as a sensor's center position and direction normal vector, calculate their intersection points and add uncertainty as hit points (global coordinates).

\item[b.] Transform the global coordinates into coordinates centered on the sensor midpoint (local coordinates) by right multiplying the obtained hit point by the DCM of the corresponding sensor.

\item[c.] Multiplying the above results with the center's global coordinates, to obtain the final simulation coordinates.

\end{itemize}

Finally, the coordinates obtained can be brought into the alignment program to verify the accuracy of the alignment results. The tracks of the flight data are chosen as the original initial trajectory, and the previous alignment result is set as the initial value of the misalignment. Then add the corrected residuals as the intrinsic resolution in order to make the simulation results more realistic. 
%The coordinates obtained by this method are faster than the GEANT4 simulation, without considering the effects of particle fragmentation, more efficient, and the accuracy is basically the same.

%\section{Analysis of simulation results}
\subsection{Method Validation}\label{subsec.stk_ana}

\begin{figure}[htbp]
\centering
\subfloat[\label{fig:no_rot_xy}]{\includegraphics{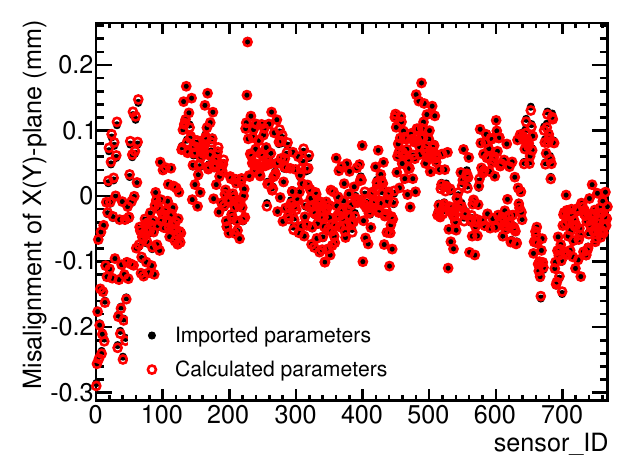}}
\subfloat[\label{fig:no_rot_z}]{\includegraphics{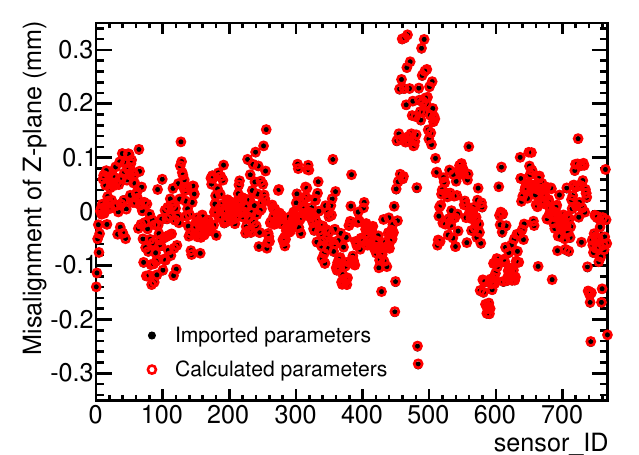}} \\
\subfloat[\label{fig:no_rot_dxy}]{\includegraphics{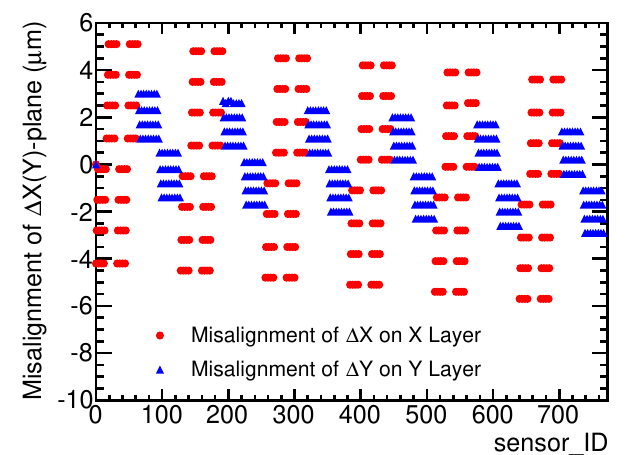}}
\subfloat[\label{fig:no_rot_dz}]{\includegraphics{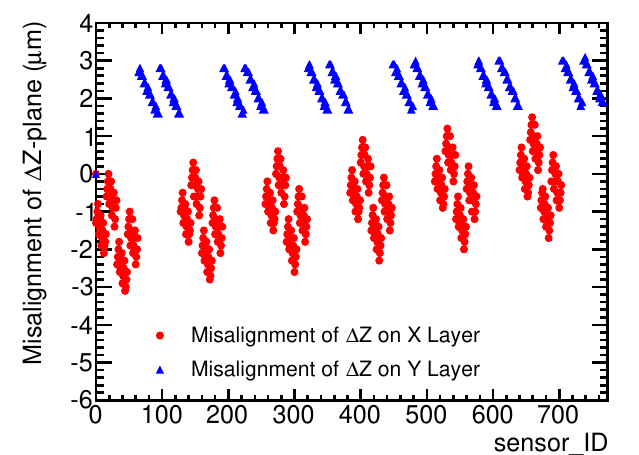}}
\caption{Variations of misalignment parameters at different sensors without rotation and uncertainty,    Figure  \ref{fig:no_rot_xy} and Figure  \ref{fig:no_rot_z} show the initial misalignment and alignment parameters distributions in the X-Y and Z plane, respectively. The difference between the aligned parameters value and the original value is shown in the Figure  \ref{fig:no_rot_dxy} and Figure  \ref{fig:no_rot_dz}.}
\label{sim_no_rot}
\end{figure}

Data without rotation and uncertainty is initially simulated to verify the viability of the alignment algorithm. Figure  \ref{sim_no_rot} depicts the initial and corrected alignment values for several sensors, with validation of simulated data demonstrating the method's effectiveness. We can see from the simulation results that our method does not exactly correct back to the original sensor location, rather there is some deflection. Only the track of STK is used in the alignment process, implying that if no other constraints are imposed, there are an infinite number of solutions satisfying the above equation. Because each layer of a single-sided silicon micro-strip can only read one-dimensional information from the particle, and since X and Y are nearly independent throughout the alignment process, the alignment findings may result in inconsistent X and Y planes, as illustrated in Figure \ref{sim_no_rot}. The rotation angle is so small in the alignment results of the flight data that converting the rotation matrix to Euler angles is problematic, so the rotation parameters are not compared.

Although each sensor's orientation follows a Gaussian distribution with a mean value in the Z plane in the simulation, the overall trend of STK is not necessarily zero. By default, this algorithm will calibrate the STK in the direction of the overall trend, which corresponds to a detector deflection, but this direction is also its own "right" orientation.

Fortunately, due to hardware restrictions, a significant sensor tilt is improbable. The distribution of the particles used for alignment is basically the same on a day-by-day basis. The more intermediate sensors pass through the most tracks, which is equivalent to selecting these sensors as the benchmark. This also ensures that the final results' deflection angle is small and does not change significantly over time.

%After confirming the algorithm's viability, the findings were checked by adding rotation and uncertainty rows to the simulated data. Figure  \ref{rot_res} shows that the correction can even go beyond the original intrinsic resolution, owing to the alignment's reliance on the residuals-based minimal Chi-square approach, which can unavoidably lead to overfitting and degeneracy with rotation parameters.

\begin{figure}[htbp]
\centering
\includegraphics{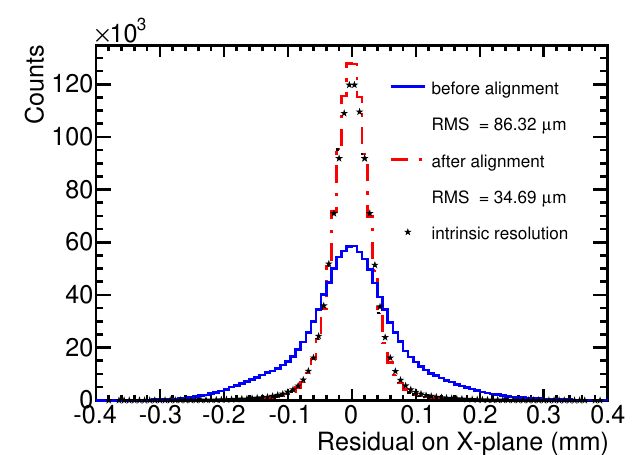}
\includegraphics{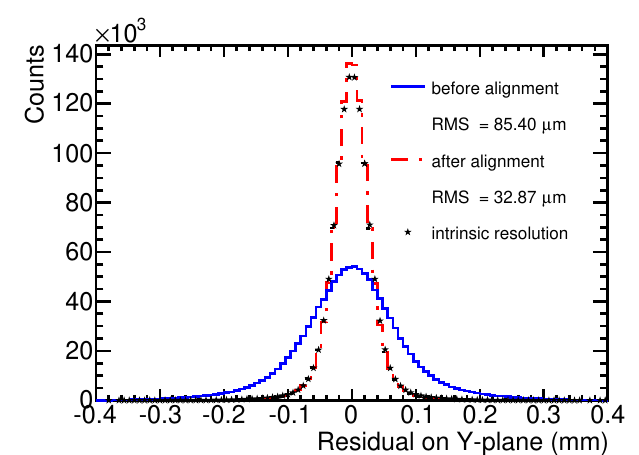}
\caption{%Track x- and y-residuals before and after alignment
The residuals of X-plane (left) and Y-plane (right). And blue line and red line present with and without alignment.}
\label{rot_res}
\end{figure}

After confirming the algorithm's viability, the results were validated by inserting rotation and uncertainty rows into the simulated data. Because the alignment depends on the minimal Chi-square method based on residuals, which can unavoidably lead to overfitting and degeneracy with rotation parameters (see Figure \ref{rot_res}), the correction can even go beyond the original intrinsic resolution.

\begin{comment}
as illustrated in Figure \ref{sim_rot}. It was also discovered that the algorithm's results could not be perfectly rectified back to the original location, and there was still a tiny deflection. However,
\begin{figure}[htb]\centering
\subfloat[\label{fig:rot_xy}]{\includegraphics{fig/rot/rot_xy.pdf}}
\subfloat[\label{fig:rot_z}]{\includegraphics{fig/rot/rot_z.pdf}} \\
\subfloat[\label{fig:rot_dxy}]{\includegraphics{fig/rot/rot_dxy.pdf}}
\subfloat[\label{fig:rot_dz}]{\includegraphics{fig/rot/rot_dz.pdf}}
\caption{Variations of misalignment parameters at different sensors with rotation and uncertainty, Figure  \ref{fig:rot_xy} and Figure  \ref{fig:rot_z} show the initial and corrected distributions of the alignment parameters in the X-Y and Z plane, respectively. The difference between the aligned parameters value and the original value is shown in the Figure  \ref{fig:rot_dxy} and Figure  \ref{fig:rot_dz}.}
\label{sim_rot}
\end{figure}
\end{comment}

\begin{figure}[ht]
\centering
\includegraphics{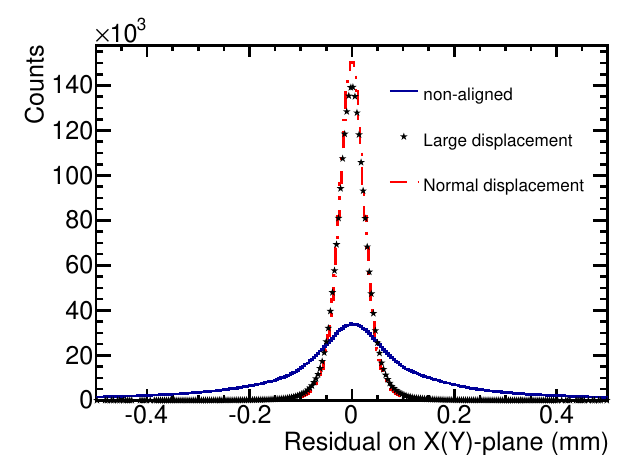}
\includegraphics{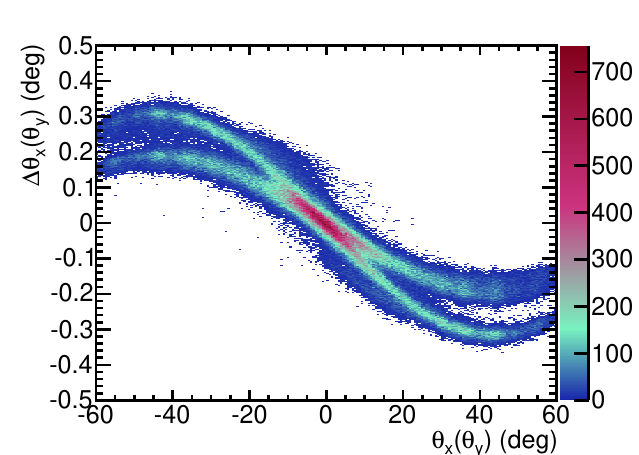}
\caption{In the case of large displacement of some layers, left panel shows the distribution of position residuals, and right panel shows the $\Delta\theta x(y)$ between the fitted track and the real track. 
As presented in the right panel, it should be roughly a straight line. 
Due to the multi-layer significant displacement, the alignment method does not correct the stretching of the tracks in the X and Y directions.}
\label{fig5}
\end{figure}

During the simulation, it was also found that if a large displacement (millimeter order) of the detector multi-layer has occurred, which is equivalent to the nominal geometric position of the multi-layer is wrong. The self-alignment could not recover to the true position, and the dispersion of the reconstructed track was larger, while the residuals remained essentially the same (shown in Figure \ref{fig5}). As a result, other sub-detectors could be combined to verify the STK alignment results.

\section{Global alignment with BGO and PSD}\label{gloabl_alignment}
\subsection{Event selection}
%The Coulomb scattering is increasing as decreasing of particle energy, particles with energies greater than 15 GeV are chosen in this work, and the number of entries selected for constructive three days is around 70,000, the statistics of such data set is enough to make result robust. Considering that the position shift is primarily sensitive to temperature during flight on orbit, DAMPE works on a  sun-synchronous orbit, where is helpful to keep a relatively stable temperature for our instrument in a short period and also beneficial to precise measurement. The daily variation of the STK's temperature is less than 1$^{\circ}$C\cite{stk_temp}, the tracks of the particles are rebuilt using Kalman filtering after alignment.

The Coulomb scattering is increasing as particle energy decreases. Particles with energies greater than 15 GeV are chosen in this work, and the number of entries selected for constructive three days is around 70,000. The statistics of such a data set is enough to make the result robust. Considering that the position shift is primarily sensitive to temperature during flight in orbit, DAMPE works in a sun-synchronous orbit, which is helpful to keep a relatively stable temperature for our instrument in a short period of time and also beneficial to precise measurement. The daily variation of the STK's temperature is less than 1$^{\circ}$C\cite{stk_temp}, the tracks of the particles are rebuilt using Kalman filtering after alignment.

\subsection{STK self-alignment}
\begin{figure}[!h]
\centering
\includegraphics{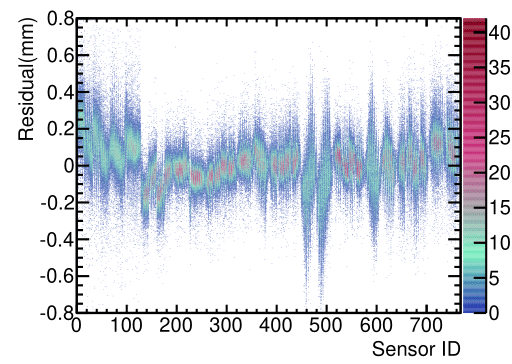}
\includegraphics{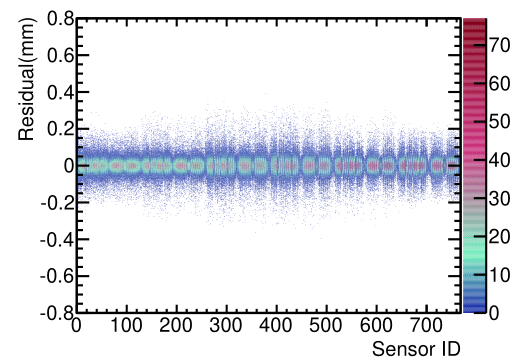}
\caption{Residuals of different sensors without (left) and with (right) alignment procedure.}
\label{fly_result}
\end{figure}

\begin{figure}[!h]
\centering
\subfloat[\label{fig:dx}]{\includegraphics[scale=0.22]{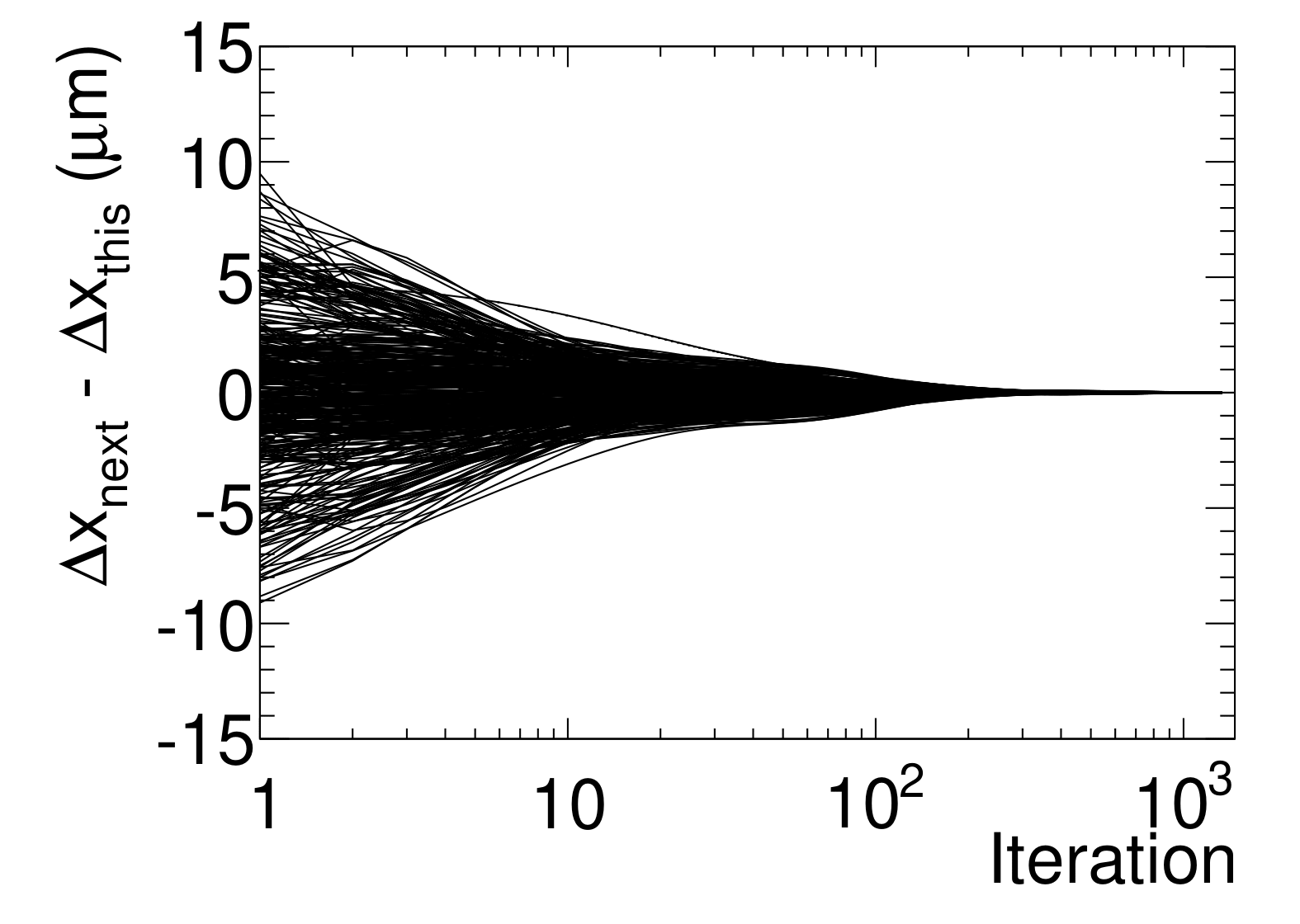}}
\subfloat[\label{fig:dy}]{\includegraphics[scale=0.22]{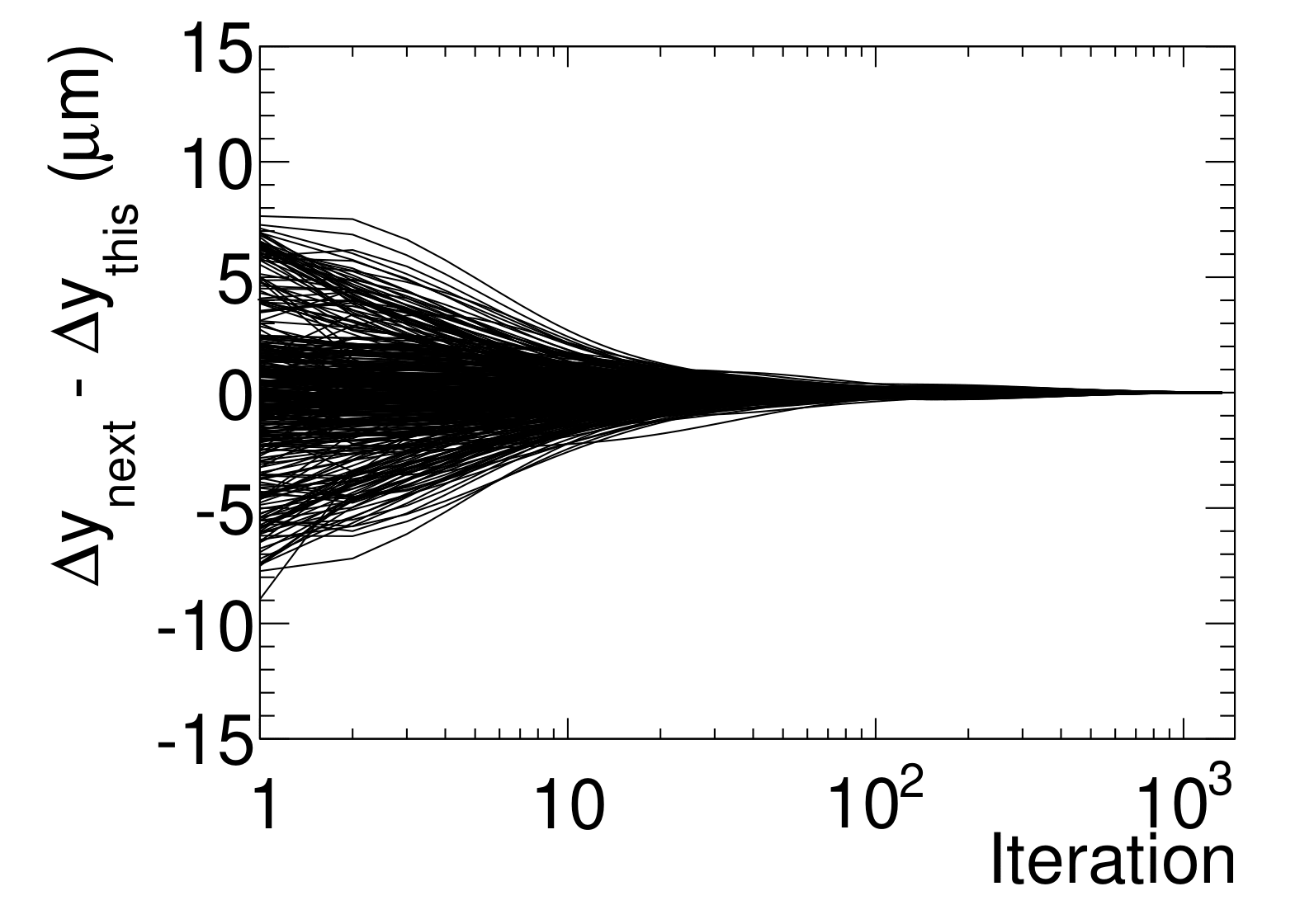}}
\subfloat[\label{fig:dz}]{\includegraphics[scale=0.22]{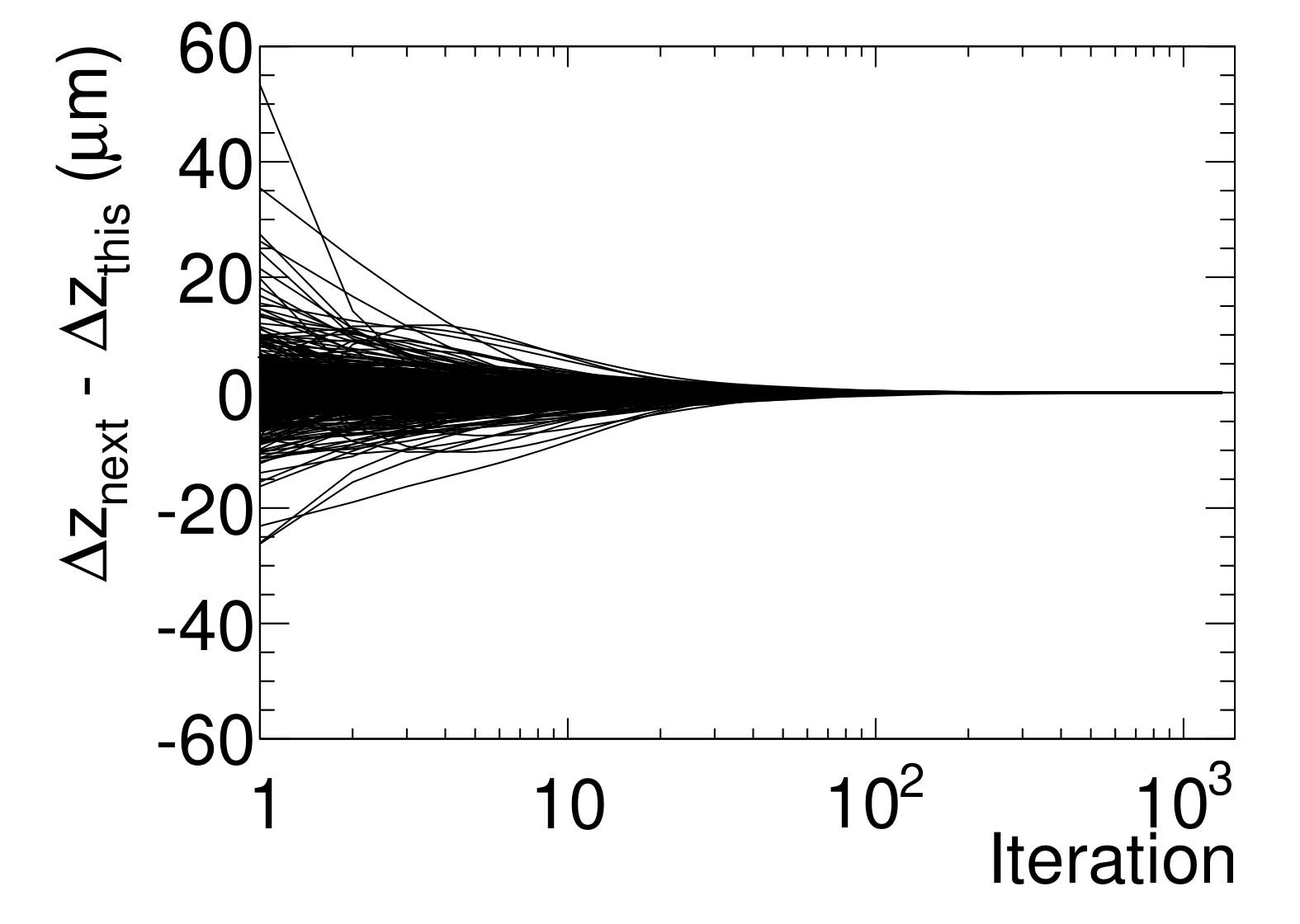}} \\
\subfloat[\label{fig:tx}]{\includegraphics[scale=0.22]{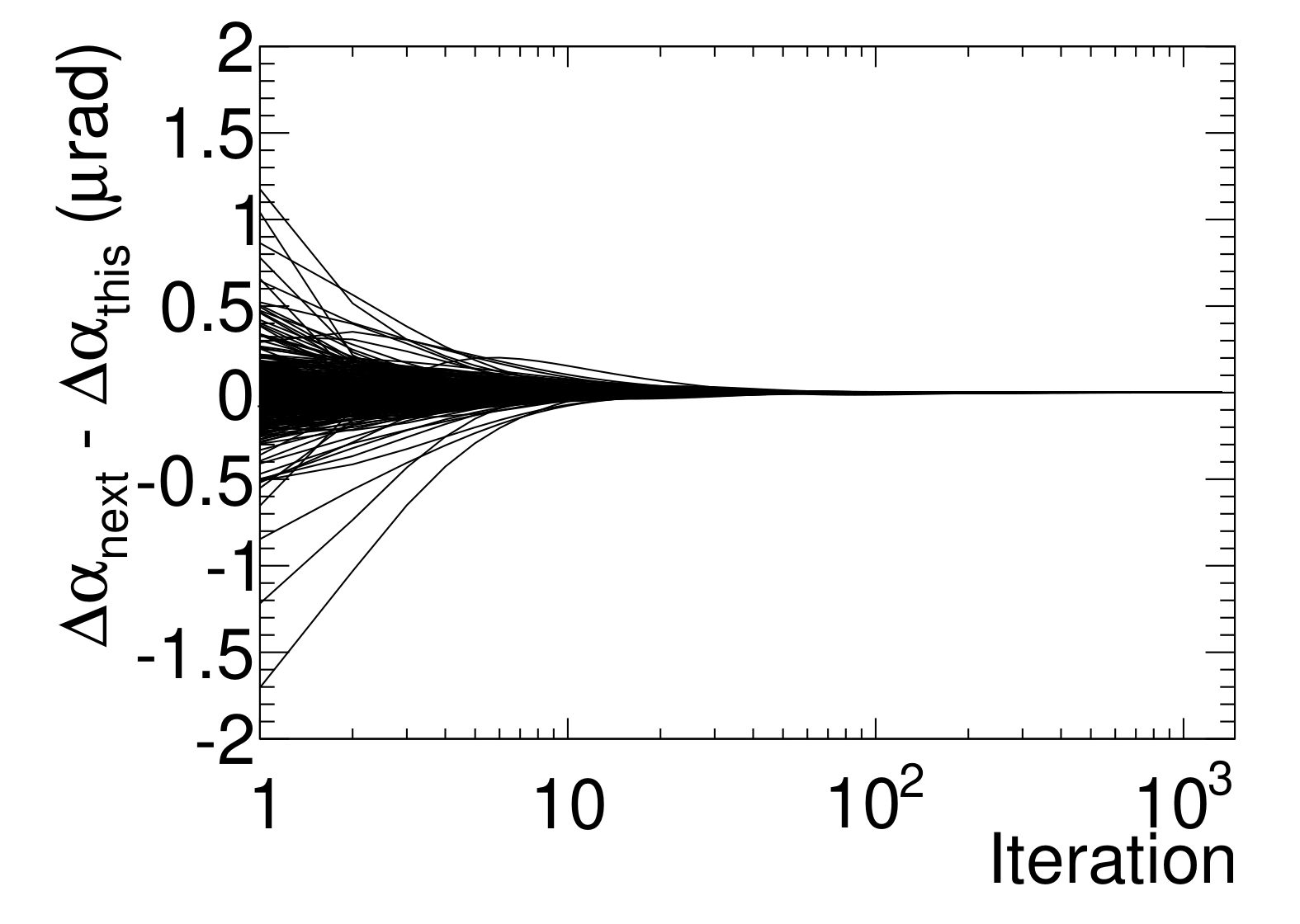}}
\subfloat[\label{fig:ty}]{\includegraphics[scale=0.22]{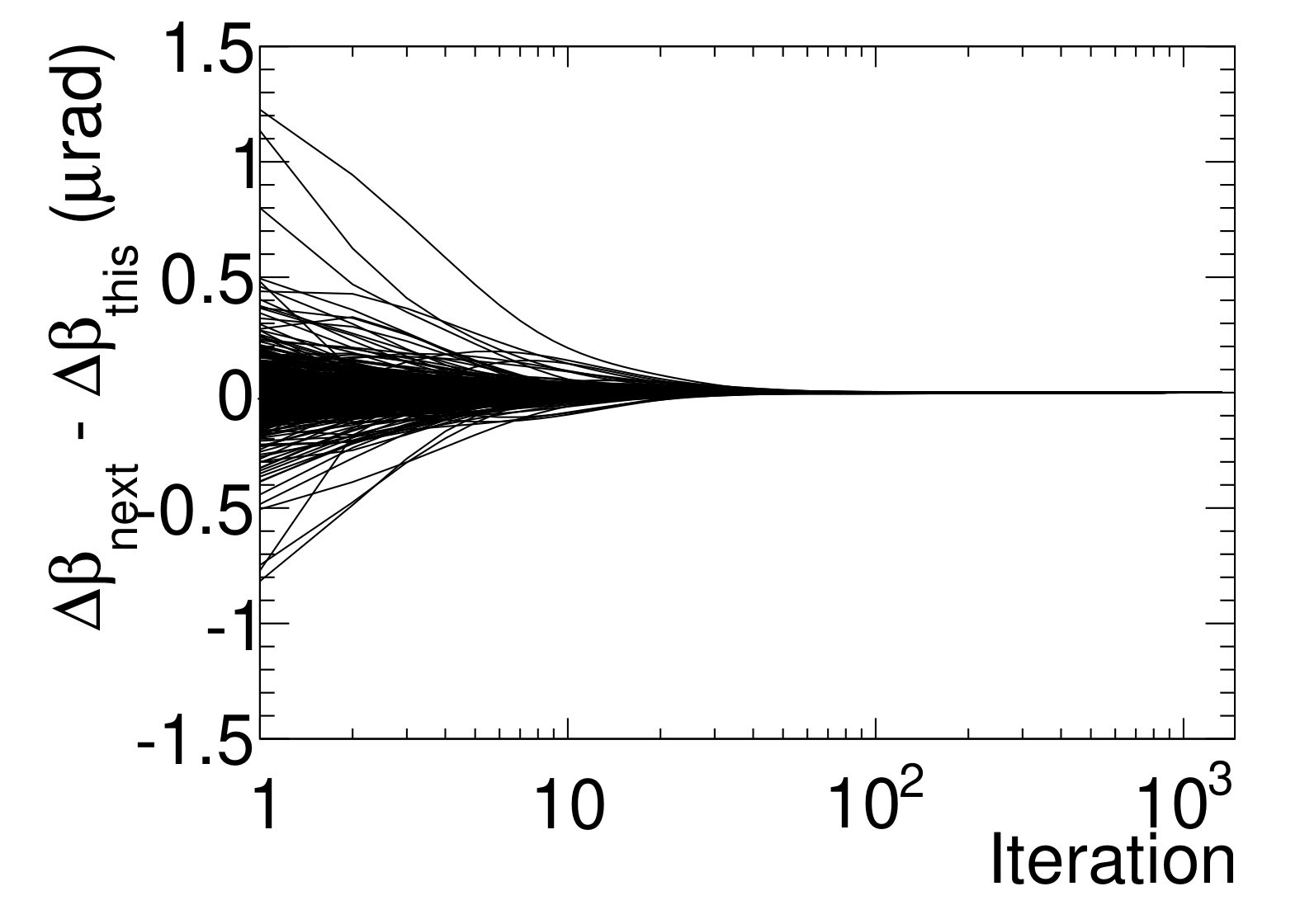}}
\subfloat[\label{fig:tz}]{\includegraphics[scale=0.22]{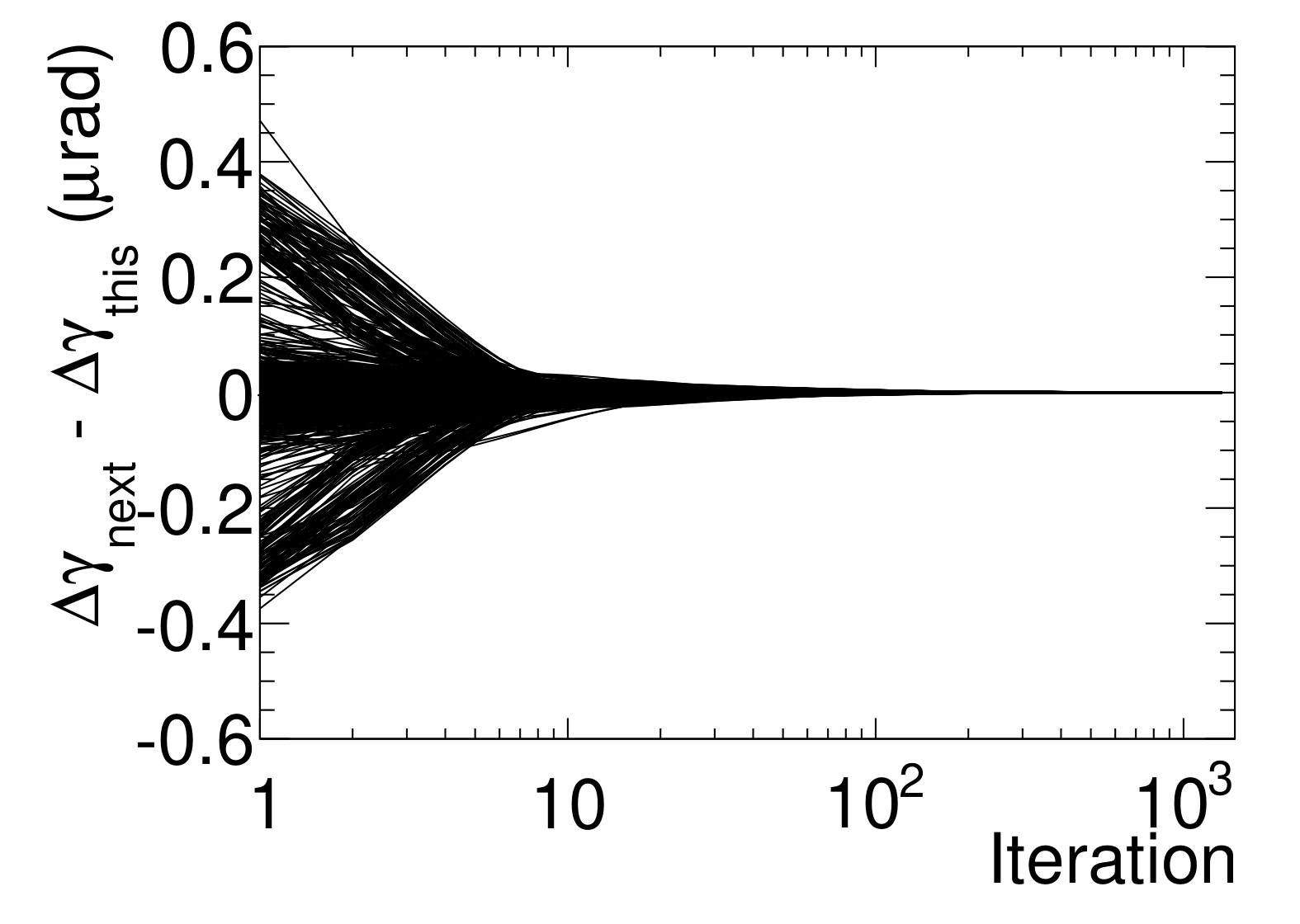}}
\caption{Convergence of variables in alignment as a function of iteration, the first row shows $\Delta x$, $\Delta y$, $\Delta z$ convergence tendencies, and the second row represents $\Delta\alpha $, $\Delta\beta$, $\Delta\gamma$, respectively.}
\label{Convergence}
\end{figure}

\begin{figure}[!h]
\centering
\includegraphics{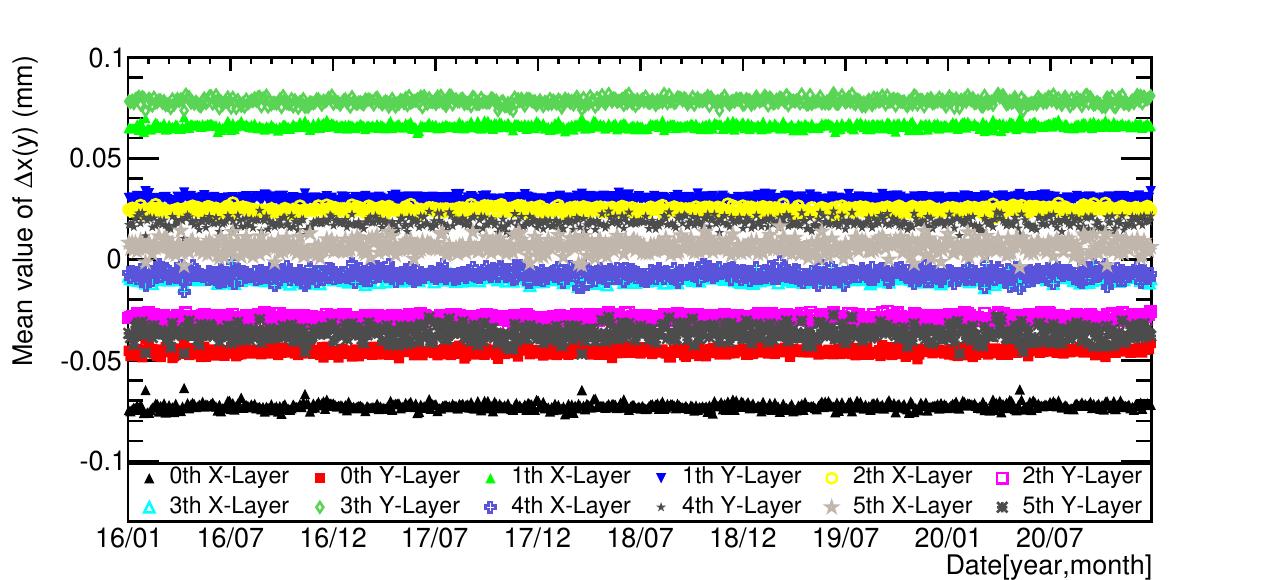}
\includegraphics{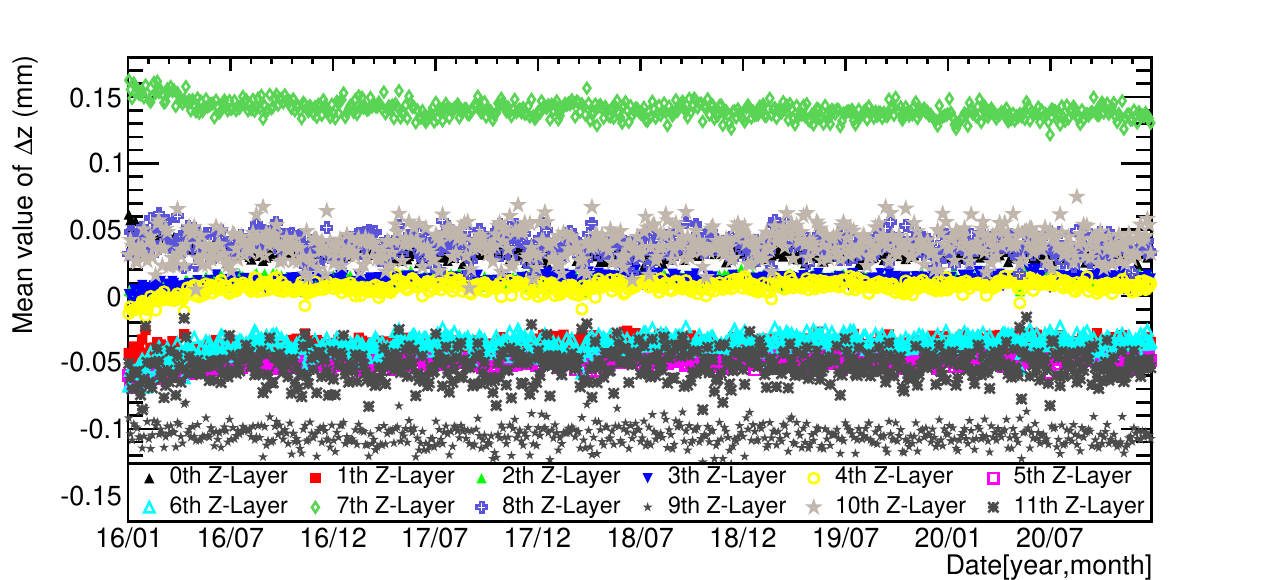}
\caption{%The variation of the $\Delta x(\Delta y), \Delta z$ or different planes as a function of observation time, a few layers deviated slightly at the start of operation and eventually reached a stable operation state.
The upper panel and bottom panel are the variation of the $\Delta x(\Delta y)$ and $\Delta z$ over time, respectively. And the status of each layers are stable, which has been show with different colors.}
\label{stabel}
\end{figure}

\begin{figure}[!h]
\centering
%\subfloat[\label{fig:vela_no}]{\includegraphics{fig/vela/vela_x.pdf}}
%\subfloat[\label{fig:vela_al}]{\includegraphics{fig/vela/vela_y.pdf}} \\
\subfloat[\label{fig:vela_1d}]{\includegraphics{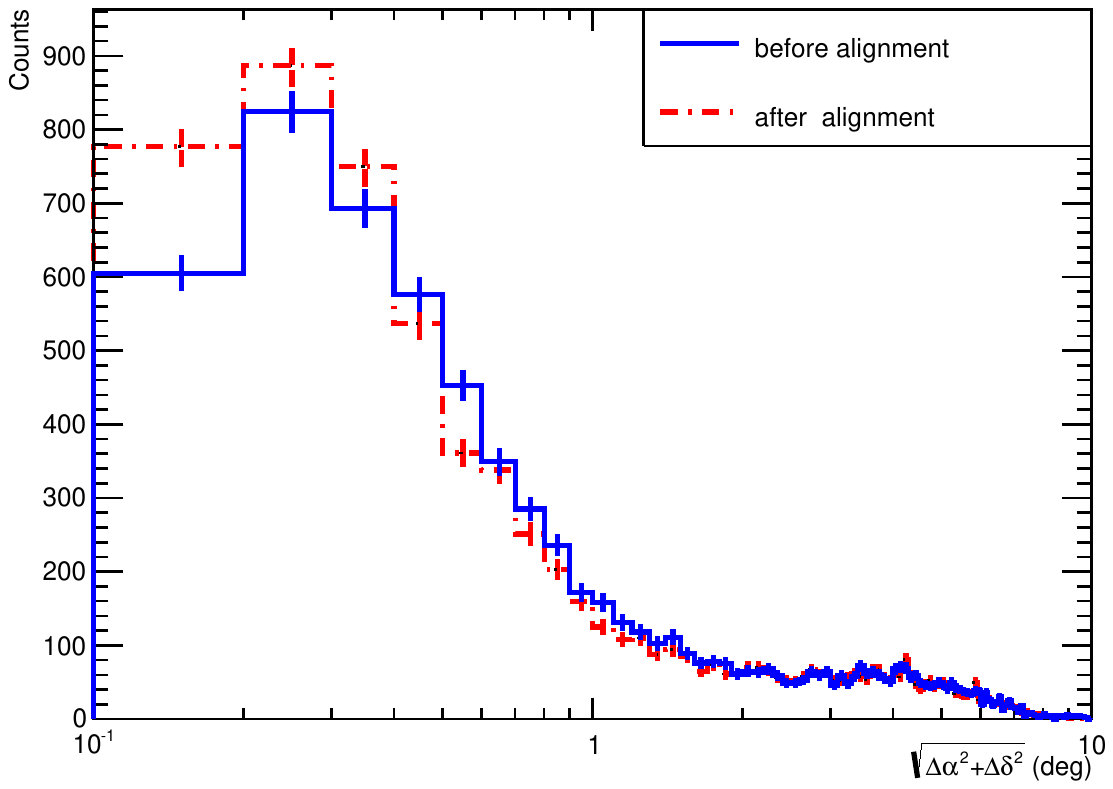}}
\caption{%Non-aligned and aligned observations of the Vela source.
The observation with (red) and without (blue) alignment of Vela source.}
\label{vela}
\end{figure}

Firstly, the data set as described in the above section is fitted with least squares to generate two linear equations in the X-Z and Y-Z planes, and then the fit points calculated from the linear equations are substituted into the alignment program to obtain a set of alignment parameters, after which the hit points are corrected using the parameters obtained, and then the revised points are re-fitted to the alignment program again, iteratively until the residuals converge (Figure \ref{Convergence}). The effect of the alignment procedure is illustrated in Figure \ref{fly_result}. Additionally, Figure \ref{stabel} show the alignment convergence and stability with time.
The results of Vela's observed entry were rebuilt in order to further evaluate the alignment's trustworthiness (see Figure  \ref{vela}).

\begin{comment}
\begin{figure}[htb]\centering
\sidesubfloat[]{\includegraphics[width=0.44\textwidth]{fig/vela/vela_no.pdf}\label{fig:vela_no}}
\hfil
\sidesubfloat[]{\includegraphics[width=0.44\textwidth]{fig/vela/vela_al.pdf}\label{fig:vela_al}}
\hfil
\sidesubfloat[]{\includegraphics[width=0.9\textwidth]{fig/vela/vela_1d.pdf}\label{fig:vela_1d}}
\caption{Non-aligned (left) and aligned (right) observations of the Vela source}
\label{vela}
\end{figure}
\end{comment}

\subsection{BGO and PSD joint alignment}\label{sec.bgo}
The STK in DAMPE plays a key role in trajectory reconstruction and defining the geometrical coordinates for DAMPE, which is crucial to almost all scientific measurement. As demonstrated in section \ref{subsec.stk_ana}, the self-alignment of STK may be biased. That is why we consider the other two sub-detectors into account to know and eliminate possible biased shift for STK itself. If the final tracks are incorrect, calibrating the PSD and BGO with the final results yields unexpected results, such as overall displacements or individual layer displacements up to the millimeter scale, which is impossible to achieve.

We know the real size and nominal position of each unit of these two sub-detectors and can forwardly calculate the path length of one incident cosmic ray particle within them using STK track information. The path length in the volume of one PSD or BGO bar is derived taking into account the shift and rotation alignment corrections. Those parameters could be imported to construct a series of functions linked with each incident particle. More details about the principle of alignment could be found in this Ref.~\cite{psd_ali}, which has been demonstrated and approved by data analysis for years and is also applicable to the BGO. The Basic principle of this alignment for PSD and BGO is to connect the ionization energy with its path length. 
% Here we import one more variable into the PSD and BGO alignment procedure compared with primary Eq. 2 \& 5 in the Ref.~\cite{psd_ali}. 
To describe the alignment of PSD and BGO, we introduce parameters as Ref.~\cite{psd_ali} (please see Sec.2  for the details):

\begin{equation}
\begin{split}
PL(\Delta x(\Delta y),\Delta z,\Delta \gamma,\Delta \beta(\Delta \alpha)) =& \frac{1}{\cos\theta} \cdot \left(a\frac{D_z}{D_{x(y)}}\Delta x(\Delta y)- a\Delta z + a\frac{D_z}{D_{x(y)}}\Delta L_i\Delta \gamma \right.\\ 
&-a\Delta L_i\Delta \beta(\Delta \alpha) -P_z-az_{c}+\frac{T}{2} \\
&\left. +a\frac{D_z}{D_{x(y)}}\left(x_{c}(y_{c})+b\frac{W}{2}-P_{x(y)}\right)\right),
\end{split}
\label{equ_psd_bgo_path}
\end{equation}
%where PL is the path length in a BGO bar's volume, $\Delta L_{i}$ is the offset along the bar of the i-th segment with respect to the bar's center, $(x_{c},y_{c},z_{c})$ is the ideal geometrical center point of one PSD bar, and T and W are the thickness and width of a BGO bar, respectively. a,b are coefficients that correspond to distinct particle incidence situations, and the track given by STK is defines by a point($P_{x},P_{y},P_{z}$) and a direction($D_{x},D_{y},D_{z}$). For each corner event, we get

where PL is the path length in the volume of a BGO bar, $\Delta L_{i}$ is the offset of the i-th segment along the bar with respect to the bar's center, $(x_c, y_c, z_c)$ is the ideal geometrical center point of one PSD bar, and T and W are the thickness and width of a BGO bar, respectively. a,b are coefficients that correspond to distinct particle incidence situations, and the track given by STK is defined by a point ($P_{x},P_{y},P_{z}$) and a direction ($D_{x},D_{y},D_{z}$). For each corner event, we get

% see the reference\cite{psd_ali} for more information.
\begin{equation}
H-\frac{D_{x(y)}}{D_{z}} V+\Delta L_{i} \Delta \gamma-\frac{\Delta L_{i}}{D_z} \Delta \alpha(\Delta \beta)=C,
\end{equation}

\begin{equation}
\begin{aligned}
C=& \frac{D_{x(y)}}{D_{z}}\left(\frac{a E_{\mathrm{dep}} \cos \theta}{S}+z_{0}+a P_{z}-a \frac{T}{2}\right) +P_{x(y)}-x_{0}\left(y_{0}\right)-b \frac{W}{2} ,
\end{aligned}
\end{equation}
where $E_{\rm dep}$ is deposited energy in one BGO bar. for all corner events, we obtain the following matrix form:

\begin{equation}
\left(\begin{matrix}
1 & \left[-\frac{D_{x(y)}}{D_z}\right]_1 & \left[\Delta L_i\right]_1 &  \left[-\frac{D_{x(y)}}{D_z}\Delta L_i\right]_1\\
1 & \left[-\frac{D_{x(y)}}{D_z}\right]_2 & \left[\Delta L_i\right]_2 & \left[-\frac{D_{x(y)}}{D_z}\Delta L_i\right]_2\\
\vdots & \vdots & \vdots & \vdots \\
1 & \left[-\frac{D_{x(y)}}{D_z}\right]_N & \left[\Delta L_i\right]_N & \left[-\frac{D_{x(y)}}{D_z}\Delta L_i\right]_N\\
\end{matrix}\right)
\left(\begin{matrix}
\Delta x(\Delta y) \\ \Delta z \\ \Delta \gamma\\
\Delta \alpha(\Delta \beta)\\
\end{matrix}\right)
=\left(\begin{matrix}
\left[C\right]_1 \\ \left[C\right]_2 \\ \vdots\\
\left[C\right]_N \\ \end{matrix}\right),
\label{equ_psd_bgo}
\end{equation}
$\Delta x(\Delta y)$, $\Delta z$, $\Delta \gamma$ and $\Delta \beta(\Delta \alpha)$ are the aligning variables for PSD and BGO that correspond to horizontal shift, vertical shift, and rotation angle in the Z and Y or X planes, respectively, and N is the number of corner events in the alignment data sample. The other specific characters correspond to the specific cases of different incident cosmic rays, which could be found in the Ref~\cite{psd_ali}. The following Figure~\ref{bgo_com} shows the change in energy loss rate of some bars of BGO before and after alignment. These bars are located in the middle of the first six layers of BGO.

\begin{figure}[!h]
\centering
\subfloat[\label{fig:dx}]{\includegraphics[scale=0.22]{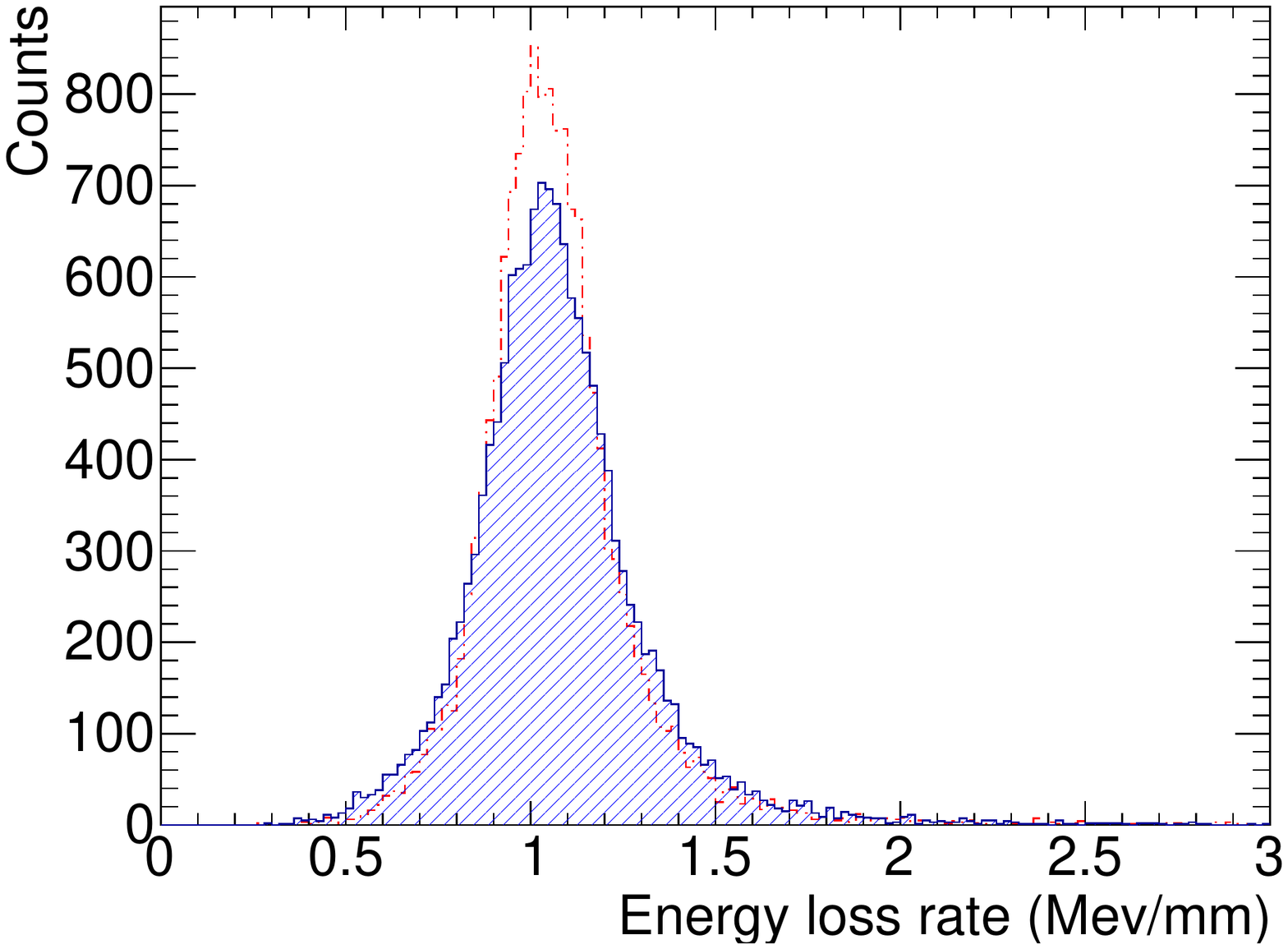}}
\subfloat[\label{fig:dx}]{\includegraphics[scale=0.22]{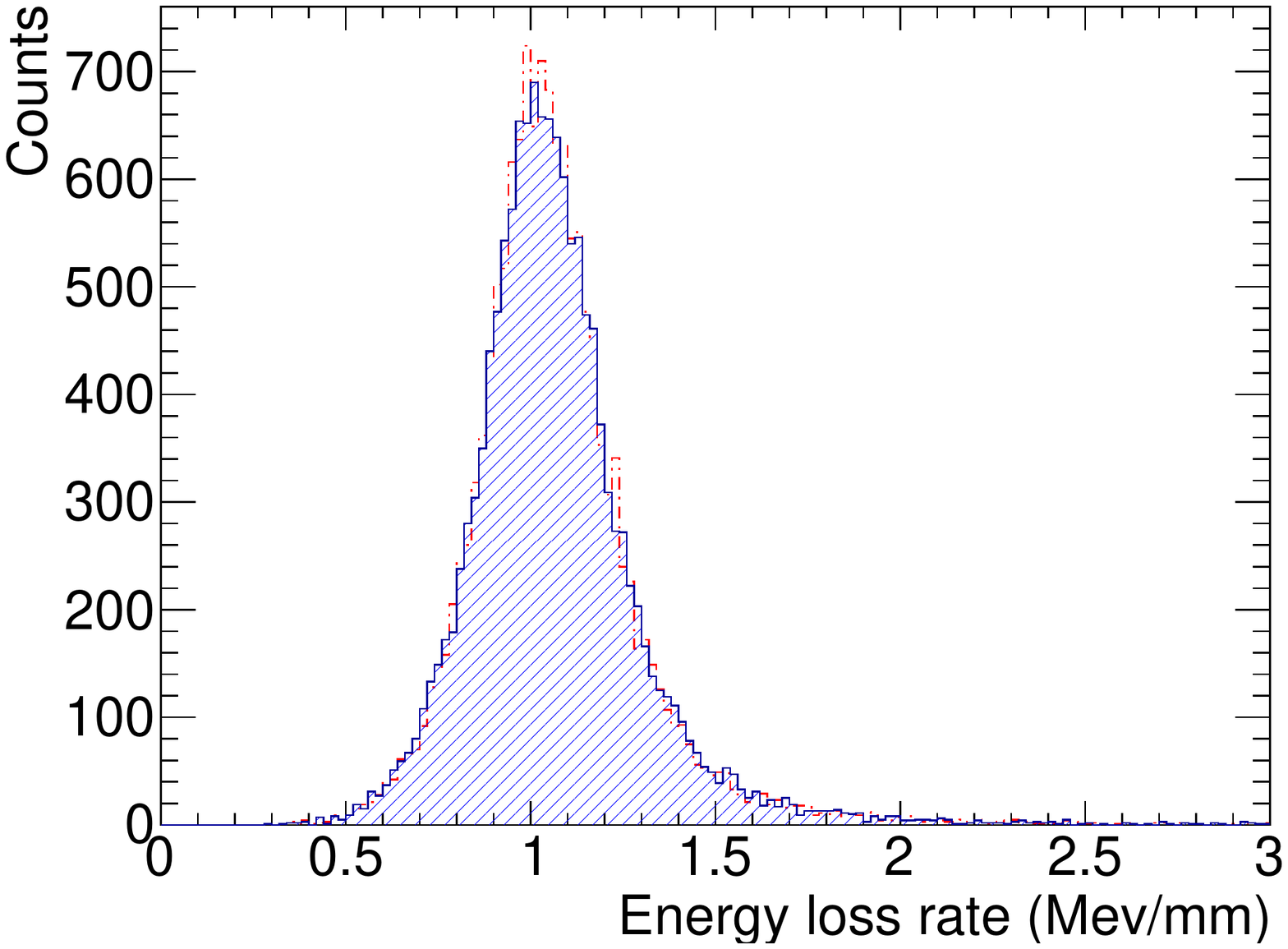}}
\subfloat[\label{fig:dx}]{\includegraphics[scale=0.22]{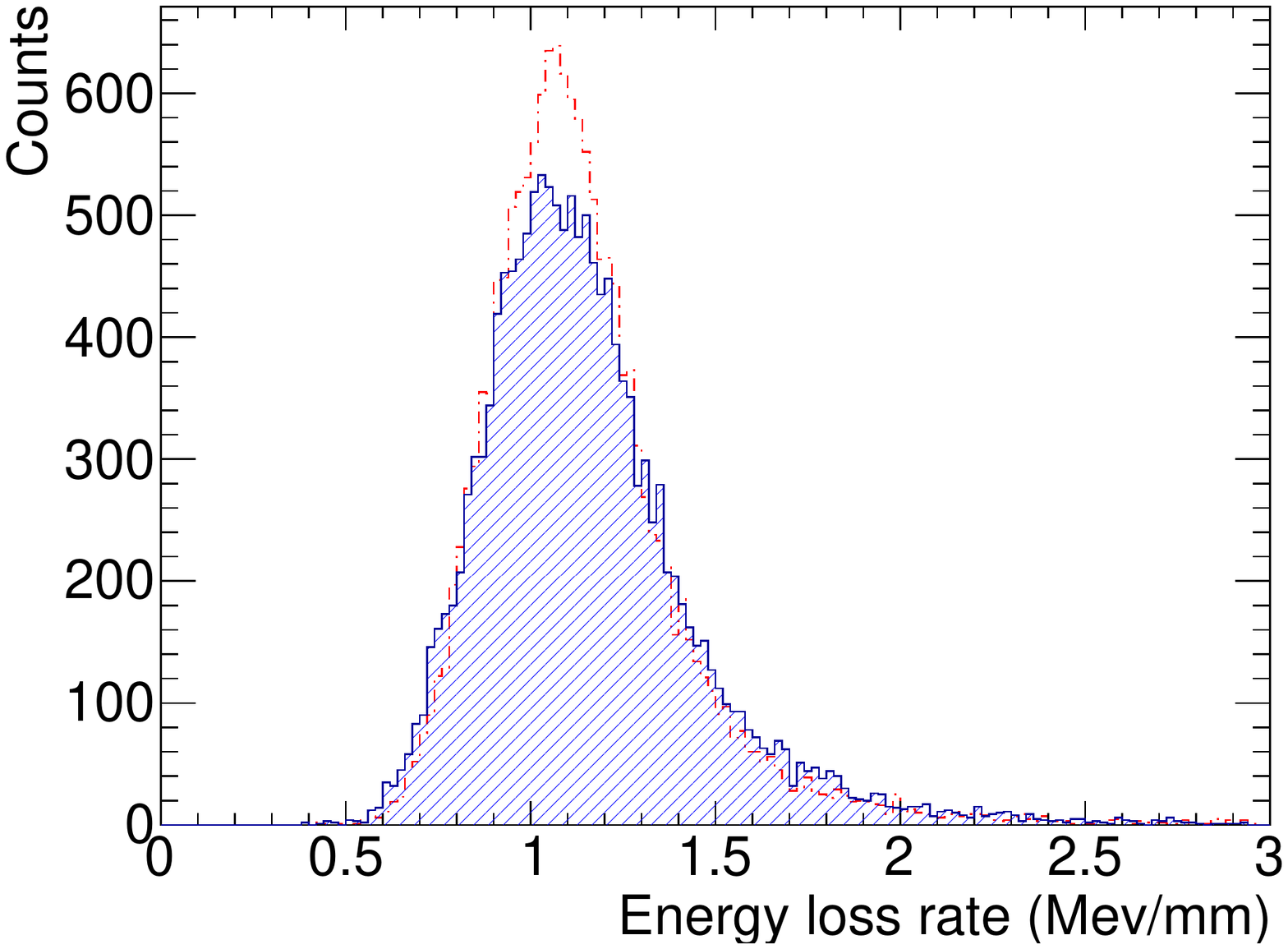}} \\
\subfloat[\label{fig:dx}]{\includegraphics[scale=0.22]{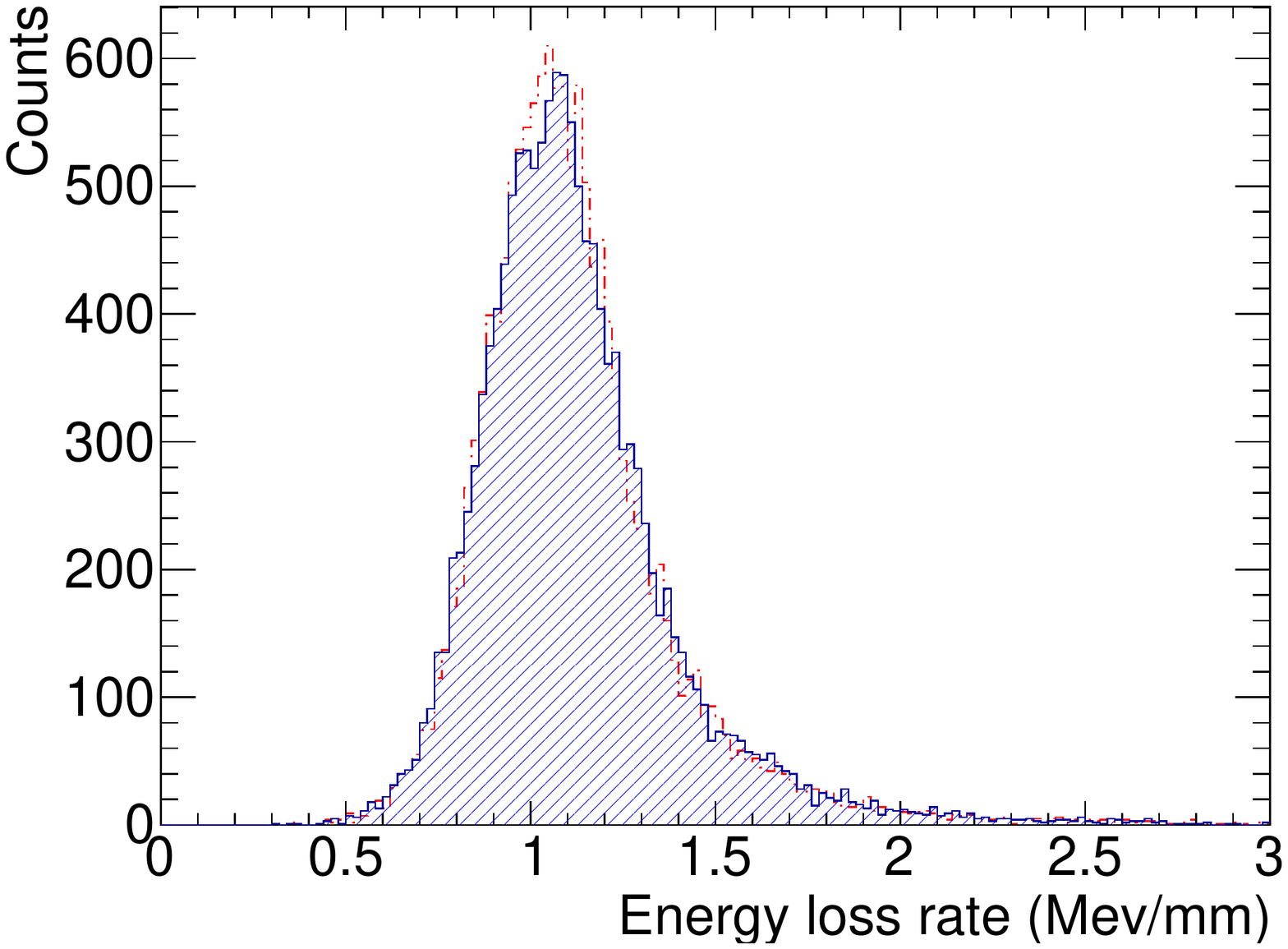}} 
\subfloat[\label{fig:dx}]{\includegraphics[scale=0.22]{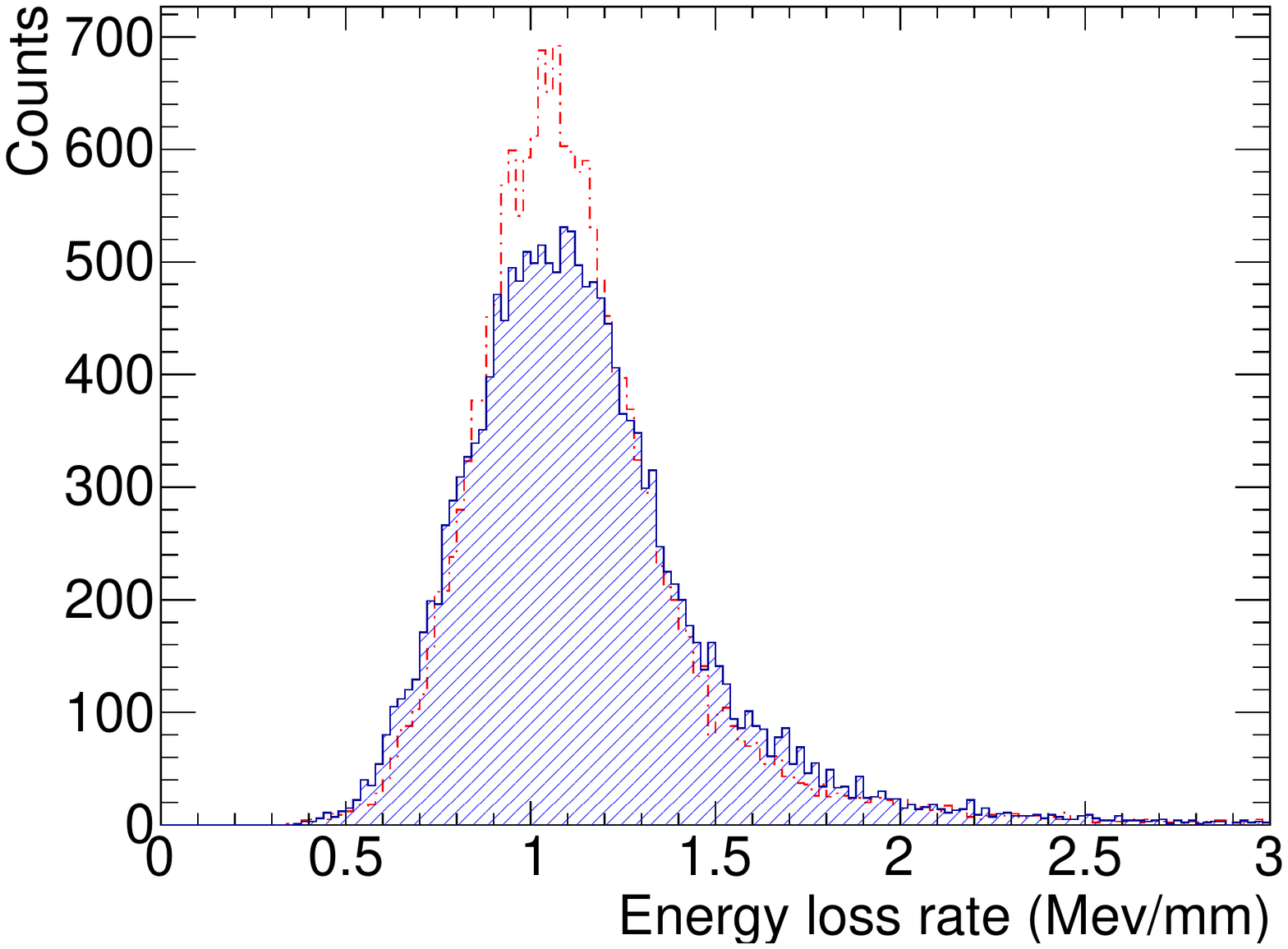}}
\subfloat[\label{fig:dx}]{\includegraphics[scale=0.22]{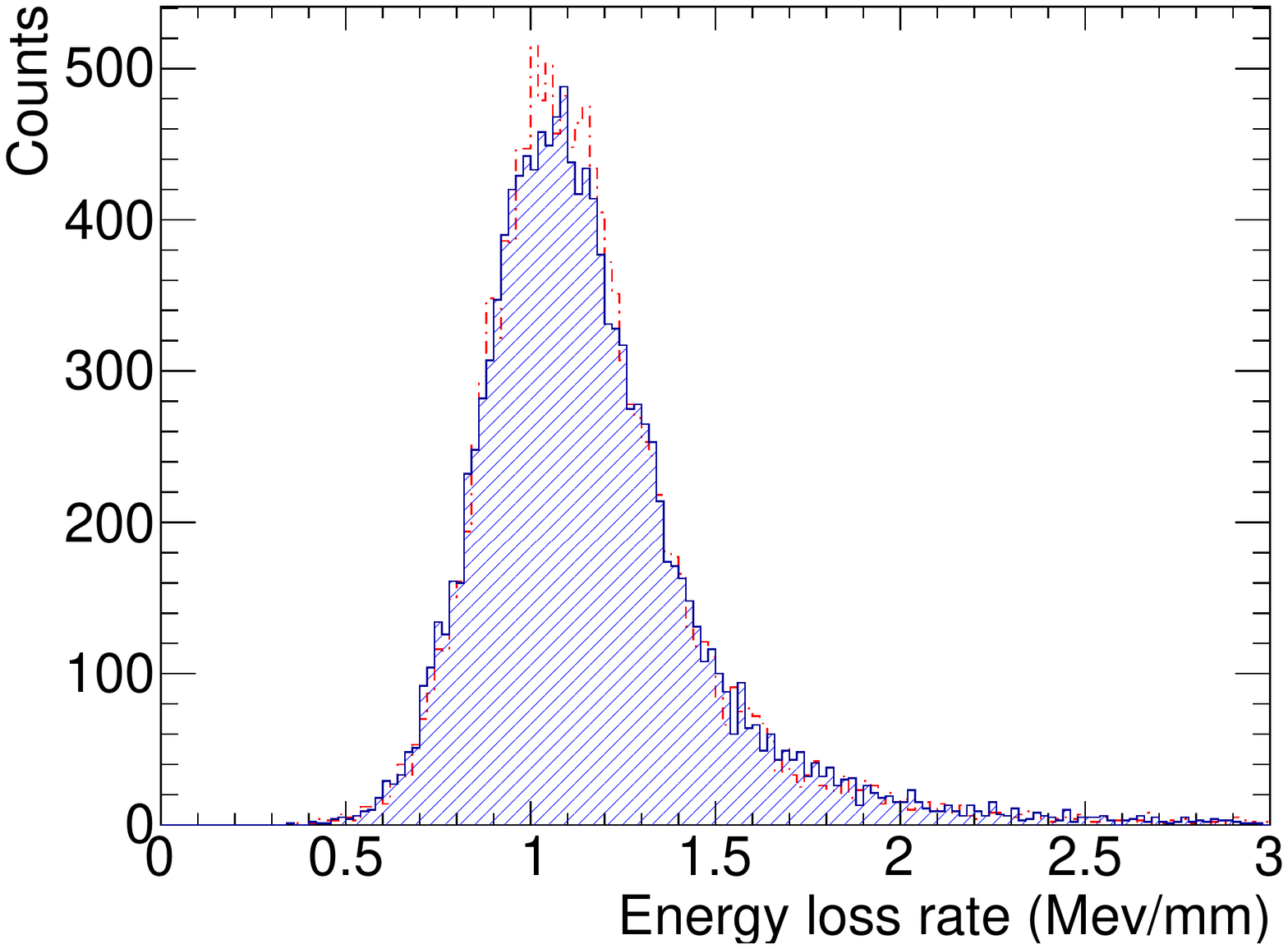}}
\caption{The energy loss rate of the bar, which is located in the first six layers of BGO. After alignment (red line), the resolution of some bars is higher than before (blue line). Some bars may have a small misalignment and have little change.}
\label{bgo_com}
\end{figure}

Due to the effect of projection and larger separation between the STK and BGO, if there is a biased global shift or rotation for the STK itself, we could judge if it is in terms of the alignment parameters of the BGO. 

\begin{itemize}
\item \emph{Rotation} \\
To begin with, a large tilt is impossible from a hardware standpoint. Nonetheless, if the alignment results are tilted, the results of the BGO alignment will appear that some bars are not horizontal. The horizontal direction of each bar is in good accord based on BGO's alignment results.

\item \emph{Displacement} \\
If the STK has an overall offset, the alignment results from BGO and PSD will indicate the same shift in the horizontal as well as an equal shift in the vertical in the same direction, and the value of the shift is the distance of the overall STK displacement.

\item \emph{Scaling} \\
As previously stated, if huge displacements exist in many layers of the STK, the final tracks may be deformed in the direction of the displacements. While it is doubtful that STK has numerous layers of considerable displacement due to its carbon fiber structure, it is more plausible that the nominal coordinates were set wrongly. The artificially multi-layer big offsets are provided in Table \ref{tabel1} to validate this effect.
\end{itemize}

\begin{table}[htbp]
	\centering
	\caption{Random displacement added to different layers (mm)}
	\label{tabel1}
	\begin{tabular}{ccccccc}
		\toprule  % 顶部线
	X & -0.45 & -0.1 & -0.15 & -0.2 & 0.15 & 0.9  \\ 
		\midrule  % 中部线
		Y & 0.3   & 0.25 & 0.19  & 0.55 & 1.3  & 2.05 \\
		\bottomrule  % 底部线
	\end{tabular}
\end{table}

\begin{table}[htbp]
	\centering
	\caption{alignment parameters of $\Delta z$ (mm) in different BGO layers.}
    \label{tabel2}
\begin{tabular}{ccc|ccc}
\hline
layer                & \multicolumn{2}{c|}{$\Delta z$}             & \multicolumn{1}{l}{Layer} & \multicolumn{2}{c}{$\Delta z$}                                 \\ \hline
\multicolumn{1}{l}{} & right & \multicolumn{1}{l|}{test} & \multicolumn{1}{l}{}      & \multicolumn{1}{l}{right} & \multicolumn{1}{l}{test} \\  \hline
x1                   & 0.74  & 0.63                       & y1                        & 0.70                      & 1.67                      \\
x2                   & 0.71  & 1.05                       & y2                        & 0.61                      & 2.28                      \\
x3                   & 0.41  & 1.74                       & y3                        & 0.47                      & 2.94                      \\
x4                   & 0.46  & 2.10                       & y4                        & 0.44                      & 3.51                      \\
x5                   & 0.34  & 2.60                       & y5                        & 0.25                      & 4.20                      \\
x6                   & 0.30  & 3.01                       & y6                        & 0.23                      & 4.83                      \\
x7                   & 0.23  & 3.89                       & y7                        & 0.22                      & 5.76                      \\ \hline
\end{tabular}
\end{table}

%According to the machining accuracy of the BGO, the error in the range of a few hundred microns is acceptable, but it is clear that $\Delta z$ grows almost linearly with the number of layers at greater displacements in Table \ref{tabel2}, which can be explained by the bias in the tracks provided by STK, Figure  \ref{wrong} shows the alignment results for the correct and test positions. In this situation, check for a large displacement of the STK layer, add the appropriate displacement to the nominal coordinates, and rerun the alignment algorithm to remove the tracks' distortion.

\begin{figure}[!h]
\centering
\subfloat[\label{fig:wr_xy}]{\includegraphics{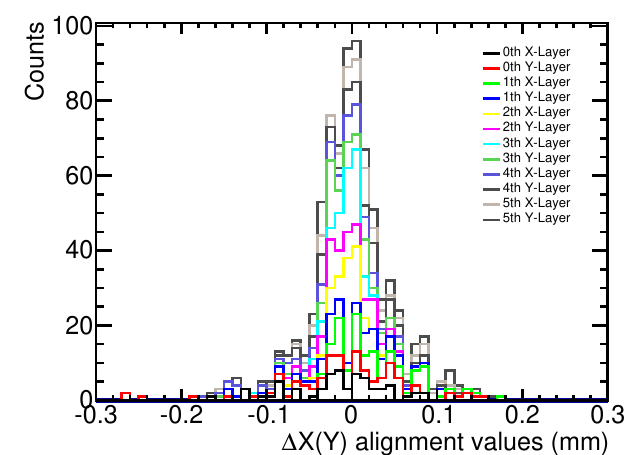}}
\subfloat[\label{fig:wr_z}]{\includegraphics{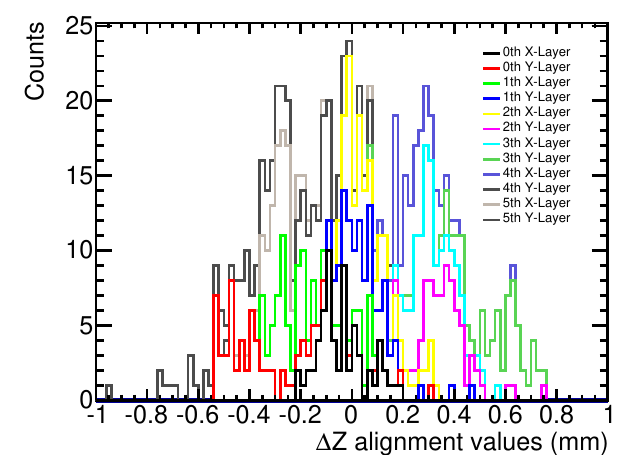}} \\
\subfloat[\label{fig:tr_xy}]{\includegraphics{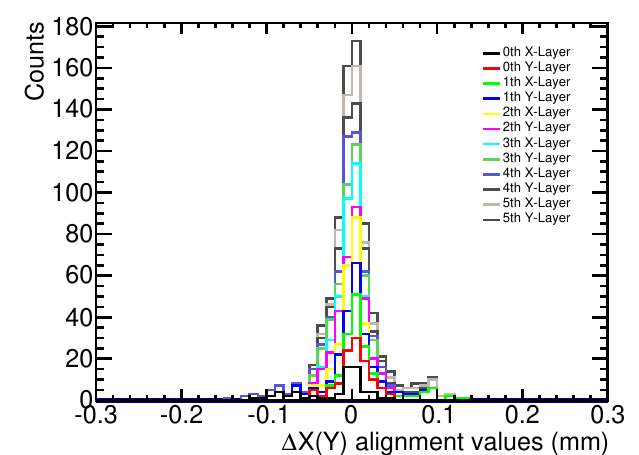}}
\subfloat[\label{fig:tr_z}]{\includegraphics{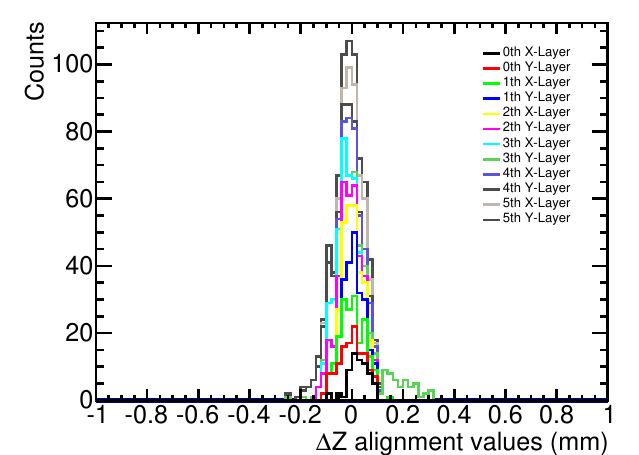}}
\caption{The distribution of $\Delta$X(Y),$\Delta$Z, the first and second rows show large displacement and no further displacement, respectively.
The first and second rows of the $\Delta$X(Y),$\Delta$Z distributions show large displacement and no further displacement, respectively.}
\label{wrong}
\end{figure}

The error in the range of a few hundred microns is acceptable according to the machining accuracy of the BGO, but it is clear that $Delta z$ grows almost linearly with the number of layers at greater displacements in Table \ref{tabel2}, which can be explained by the bias in the tracks provided by STK. Figure \ref{wrong} shows the alignment results for the correct and test positions. In this situation, check for a large displacement of the STK layer, add the appropriate displacement to the nominal coordinates, and rerun the alignment algorithm to remove the tracks' distortion. The detrimental consequences of STK self-alignment can thus be verified and addressed in this way.

\section{Discussion and conclusion}\label{sec:conclusion}

%In this work, we have presented results of the DAMPE global alignment with cosmic-ray tracks. Firstly we consider the STK self-alignment with a large number of high energy incident particles, then make a joint analysis with BGO. After corrected by the global alignment, the flight data are well consistent with the simulations.

In this analysis, we firstly carry out the STK self-alignment with a large number of incident charge particles, then make a joint analysis with PSD and BGO to fulfill the global alignment of DAMPE. After this correction, the inner misalignments of STK, PSD and BGO, and the displacement among them are well eliminated. As a result, the flight data show a very good consistency with the MC simulations.

%As the alignments of PSD and BGO primarily depend on the tracks reconstructed by STK, a rotation matrix is applied to generate pseudo simulated events to evaluate the potential uncertainty of the STK self-alignment \mkgreen{cyue: As a reader, I do not understand the relation between the former part and the latter one}. 
To evaluate the potential uncertainty of the STK self-alignment, we generate a large mount of numerical simulations by assuming a misalignment with a rotation matrix. The STK self-alignment achieves a good residual convergence even if we apply a rotation matrix with very large displacements of STK strips. However, in that case, the reconstructed track after the alignment correction shows a systematical deviation from the incident direction. Also, the BGO alignment base on those events with biased tracks show unexpected large deviations (see Sec.\ref{sec.bgo}). So a large displacement of the STK strip(s) cannot be easily identified from the STK self-alignment procedure. Thereby, the investigation of the unit displacements  in PSD and BGO is necessary to ascertain whether the STK self-alignment correction is reasonable. This is extremely significant as the alignments of PSD and BGO heavily rely on the tracks reconstructed by STK.

After the STK self-alignment, the $\rm RMS_{99.73\%}$ of the STK position residuals in flight data is about 37.26 $\rm \mu m$. By applying the alignment procedure for the STK tracks, the energy resolution of BGO is significantly improved, and the distribution of STK and BGO alignment parameters is also within a reasonable range. A more precise track reconstruction can be achieved via the global alignment, which is crucial to meet the high accuracy requirement of DAMPE.

\section*{Acknowledgements}
This work was funded by the National Natural Science Foundation of China (Grant Nos. U1831206, 11921003, 12103094, 12073087), Scientific Instrument Developing Project of the Chinese Academy of Sciences, Grant No. GJJSTD20210009, and the Natural Science Foundation of Jiangsu Province (No. BK20201107).



%\nocite{*}
\bibliography{reference}
\bibliographystyle{elsarticle-num}



%\end{thebibliography}
\end{document}